\documentclass{aa}  
\bibpunct{(}{)}{;}{a}{}{,} 
\usepackage[utf8]{inputenc}
\usepackage{txfonts}
\usepackage{url}
\usepackage{enumitem}

\usepackage{amssymb,amsmath,natbib,enumitem,longtable,psfig}
\usepackage{graphicx}
\def\gsim{ \lower .75ex \hbox{$\sim$} \llap{\raise .27ex \hbox{$>$}} }
\def\lsim{ \lower .75ex\hbox{$\sim$} \llap{\raise .27ex \hbox{$<$}} }

\def\beq{\begin{equation}}
\def\eeq{\end{equation}}

\def\sw{{\it Swift}}
\def\fe{{\it Fermi}}

\def\ep{$E_{\rm p}$}

\def\liso{$L_{\rm iso}$}
\def\eiso{$E_{\rm iso}$}
\def\ama{$E_{\rm p}-E_{\rm iso}$}
\def\yone{$E_{\rm p}-L_{\rm iso}$}
\def\ghi{$E_{\rm p}-E_{\gamma}$}

\def\th{$\theta_{\rm jet}$}

\def\tp{$t_{\rm p}$}
\def\tpg{$T_{\rm p,\gamma}$}
\def\tpul{$t_{\rm p}^{\rm UL}$}
\def\tpll{$t_{\rm p}^{\rm LL}$}
\def\tdur{$T_{90}$}
\def\tph{$T_{\rm ph}$}

\def\G{$\Gamma_{0}$}
\def\GLL{$\Gamma_{0}^{LL}$}
\def\GUL{$\Gamma_{0}^{UL}$}

\def\lisocom{$L'_{\rm iso}$}
\def\eisocom{$E'_{\rm iso}$}

\def\epcom{$E'_{\rm p}$}

\usepackage[normalem]{ulem}

\begin{document}

\title{Bulk Lorentz factors of Gamma-Ray Bursts}
\titlerunning{Bulk Lorentz factor of GRBs}
\authorrunning{G. Ghirlanda et al.}
\author{G. Ghirlanda\inst{1,2}\thanks{E--mail:giancarlo.ghirlanda@brera.inaf.it},  F. Nappo\inst{3,1}, G. Ghisellini\inst{1}, A. Melandri\inst{1}, G. Marcarini\inst{2},\\ L. Nava\inst{1,2}, O. S. Salafia\inst{2,1}, S. Campana\inst{1}, R. Salvaterra\inst{4}. \\ }
\institute{$^{1}$INAF -- Osservatorio Astronomico di Brera, via E. Bianchi 46, I-23807 Merate, Italy.\\
$^{2}$Dipartimento di Fisica G. Occhialini, Universit\`a di Milano Bicocca, Piazza della Scienza 3, I-20126 Milano, Italy.\\
$^{3}$Universit\`a degli Studi dell'Insubria, via Valleggio 11, I-22100 Como, Italy.\\
$^{4}$INAF -- IASF Milano, via E. Bassini 15, I-20133 Milano, Italy. \\
$^{5}$INAF -- Osservatorio Astronomico di Trieste, via  G.B. Tiepolo, 11, I--34143 Trieste, Italy.\\
}

\date{}


\abstract{
Knowledge of the bulk Lorentz factor \G\ of gamma-ray bursts (GRBs) allows us to compute their comoving frame properties shedding light on their physics. 
Upon collisions with the circumburst matter, the fireball of a GRB 
starts to decelerate, producing a peak or a break (depending on the circumburst density profile) in the light curve of the afterglow. 
Considering all bursts with known redshift and with an early coverage of their emission, we find 67 GRBs (including one short event) with a peak in their optical or GeV light curves at a time \tp. For another 106 GRBs we set an upper limit \tpul. The measure of \tp\ provides the bulk Lorentz factor \G\ of the fireball before deceleration. We show that \tp\ is due to the dynamics of the fireball deceleration and not to the passage of a characteristic frequency of the synchrotron spectrum across the optical band. 
Considering the \tp\  of  66 long GRBs and the 85 most constraining upper limits, we  estimate  \G\ or a lower limit  \GLL.  Using censored data analysis methods, we reconstruct the most likely distribution of \tp.  All \tp\ are larger than the time \tpg\ when the prompt $\gamma$--ray emission peaks, and are much larger than the time \tph\ when the fireball becomes transparent, that is, \tp$>$\tpg$>$\tph. The reconstructed distribution of \G\ has median value $\sim$300 (150) for a uniform (wind) circumburst density profile. In the comoving frame, long GRBs have typical isotropic energy, luminosity, and peak energy  $\langle E_{\rm iso}\rangle = 3 (8) \times 10^{50}$ erg, $\langle L_{\rm iso}\rangle = 3 (15) \times 10^{47}$ erg s$^{-1}$ , and $\langle E_{\rm peak}\rangle = 1(2) $ keV in the homogeneous (wind) case.
We confirm that the significant correlations between \G\ and the rest frame isotropic energy (\eiso), luminosity (\liso) and peak energy (\ep) are not due to selection effects. When combined, they lead to the observed \ama\ and \yone\ correlations. Finally, assuming a typical opening angle of 5 degrees, we derive the distribution of the jet baryon loading  which is centered around a few $10^{-6} {\rm M_{\odot}}$. }

\maketitle

\section{Introduction}

The relativistic nature of Gamma-Ray Bursts (GRBs) was originally posed on theoretical grounds 
\citep{Goodman:1986lr,Paczynski:1986fk,Krolik:1991qy,Fenimore:1993uq,Baring:1997fj,Lithwick:2001kx}: the small size of the emitting region, as implied by the observed millisecond variability, 
would make the source opaque due to $\gamma$--$\gamma$ pair production unless it expands 
with bulk Lorentz factor \G$\sim$100--1000 \citep[e.g.][]{Piran:1999yq}.
Refinements of this argument applied to specific GRBs  \citep{Abdo:2009rt,Abdo:2009vn,Ackermann:2010ys,Hascoet:2012fr,Zhao:2011zr,Zou:2010ul,Zou:2011ve,Tang:2015lr} led to some estimates of \G. 
A direct confirmation that GRB outflows are relativistic was found in 970508 \citep{Frail:1997pd}: 
the suppression of the observed radio variability ascribed to scintillation induced by Galactic dust 
provided an estimate of the source relativistic expansion. 
Similarly, the long-term monitoring in the radio band of GRB 030329 
\citep{Pihlstrom:2007dq,Taylor:2004bh,Taylor:2005lq}
allowed to set limits on the expansion rate \citep{Mesler:2012rr}.

In the standard fireball scenario, after an optically thick acceleration phase the ejecta 
coast with constant bulk Lorentz factor \G\ before decelerating due to the interaction 
with the external medium. 
\G\ represents the maximum value attained by the outflow during this dynamical evolution. 
Direct estimates of \G\ became possible in the last decade 
thanks to the early follow up of the afterglow emission. 
The detection of an early afterglow peak, \tp$\sim$150--200 s, in the NIR light curve of GRB
060418 and GRB 060607A provided one of the first estimates of \G\ 
\citep{Molinari:2007fj}.

The recent development of networks of robotic telescopes 
(ROTSE--III: \citealt{Akerlof:2003wd};
GROCSE: \citealt{Park:1997nx};
TAROT: \citealt{Klotz:2009eu}; 
SkyNet: \citealt{Graff:2014oq};
WIDGET: \citealt{Urata:2011qe};
MASTER: \citealt{Lipunov:2004kl};
Pi of the Sky: \citealt{Burd:2005ai};
RAPTOR: \citealt{Vestrand:2002tg};
REM: \citealt{Zerbi:2001dp}; 
Watcher: \citealt{Ferrero:2010hc})
has allowed us to follow up the early optical emission of GRBs. 
Systematic studies \citep[][G12 hereafter]{Liang:2010lr,Lu:2012bs,Ghirlanda:2012th}
derived the distribution of \G\ and its possible correlation with other observables. 
G12 found:
\begin{enumerate}
\item that different distributions of  \G\ are obtained according to the density profile of the circumburst medium;  
\item the existence of a correlation  \G$^2\propto L_{\rm iso}$ (tighter with respect to that with \eiso); 
\item the presence of a linear correlation \G\ $\propto$ \ep.
\end{enumerate}
As proposed by G12, the combination of these correlations provides a possible interpretation of the spectral 
energy correlations \ama\  \citep{Amati:2002fy} and \yone\ \citep{Yonetoku:2004fv} as the result of larger  
\G\ in bursts with larger luminosity/energy and peak energy.  Possible interpretations of the correlations between \G\ and the GRB luminosity have been proposed in the context of neutrino or magneto--rotation powered jets \citep{Lu:2012bs,Lei:2013hh}. G12 \cite[see also][]{Ghirlanda:2013la} showed that a possible relation between \G\ and the jet opening angle \th\ could also justify the \ghi\ correlation (here $E_\gamma$ is the collimation corrected energy). 

In order to estimate \G, we need to measure the onset time \tp\ of the afterglow. 
If the circumburst medium is homogeneous, this is revealed by an early peak in the light curve,
corresponding to the passage from the coasting to the deceleration phase of the fireball.
On the other hand, if a density gradient due to the progenitor wind is present, the bolometric 
light curve is constant until the onset time. However, also in the wind case, a peak could be observed if pair production ahead of the 
fireball is a relevant effect as discussed in G12, for example.

G12 considered 28 GRBs with a clear peak in their optical light curve, 
and included three GRBs with a peak in their GeV light curves 
(as observed by the Large Area Telescope  -- LAT -- on board \fe). 
Early \tp\ measurements are limited by the 
time needed to start the follow--up observations.
The LAT (0.1--100 GeV), with its large field of view, performs observations simultaneously
to the GRB prompt emission for GRBs happening within its field of view. 
The detection of an early peak in the GeV light curve, if interpreted as afterglow from the 
forward shock \citep[e.g.][]{Ghisellini:2010lq,Kumar:2010dz}, provides 
the estimate of the earliest \tp\ (i.e. corresponding to the largest \G). 
In the  short GRB 090510, the LAT light curve peaks at $\sim$0.2 s corresponding to 
\G$\sim$ 2000 \citep{Ghirlanda:2010pb,Ackermann:2010fu}. Recently it has been shown that 
upper limits on \G\ can be derived from the non-detection of GRBs by the LAT \citep{Nava:2017kl} and that 
such limits are consistent with lower limits and detections reported in the literature. 

Precise and fast localisations of GRB counterparts, routinely performed by \sw, coupled to 
efficient follow up by robotic telescope networks, allowed us  to follow the optical emission starting relatively soon after the GRB trigger.
However, a delay of a few hundred seconds can also induce a bias against the measure of early--intermediate \tp\ values 
\citep{Hascoet:2012fr}. As argued by \cite{Hascoet:2012fr}, the distribution of \G, derived through measured \tp, could lack intermediate to large values of \G\ (corresponding to intermediate to early values of \tp) and the \G--\eiso\ correlation could be a boundary, missing several bursts with large \G\ (i.e. because of the lack of early \tp\ measurements). 

For this reason, upper limits are essential to derive the distribution  of \G\ in GRBs and its 
possible correlation with other prompt emission properties (\eiso, \liso, \ep) and, 
in general, to study the comoving frame properties of the population. 
To this aim, in this paper 
(i) we collect the available bursts with an optical \tp,  
expanding and revising the previously published samples, and 
(ii) we collect a sample of bursts with upper limits on \tp. 
Through this censored data sample we reconstruct the distribution of \G\ accounting 
(for the first time) for upper limits. 
We then employ Monte Carlo methods to study the correlations between \G\ 
and the rest frame isotropic energy/luminosity and peak energy.   

The sample selection and its properties are presented in \S2, \S3, and \S4, respectively. 
The different formulae for the estimate of the bulk Lorentz factor \G\ appearing in 
the literature are presented and compared in \S5. 
In \S6 the distribution of \G\ and its correlation (\S7) with \eiso, \liso\ and \ep\ are studied. 
Discussion and conclusions follow in \S8. 
We assume a flat cosmology with $h=\Omega_\Lambda=0.7$.

\section{The sample}

We consider GRBs with measured redshift $z$ and well constrained spectral parameters of 
the prompt emission. 
For these events it is possible to estimate the isotropic energy \eiso\ and luminosity 
\liso\ and the rest frame peak spectral energy \ep\  (i.e. the peak of the $\nu F_{\nu}$ spectrum). 

\G\ can be estimated from the measure of the peak \tp\ of the afterglow light curve interpreted 
as due to the deceleration of the fireball. 
We found in the literature 67 GRBs (66 long and 1 short) with an estimate of \tp\ (see Tab. \ref{tabTOT}). 
Of these, 59 \tp\ are obtained from the optical and 8 from the  GeV light curves. 
Through a systematic search of the literature we collected 106 long GRBs whose optical 
light curve, within one day of the trigger, decays with no apparent \tp. 
These GRBs provide upper limits \tpul. 
Details of the sample selection are reported in the following sections. 

\subsection{Afterglow onset \tp}

G12 studied a sample of 30 long GRBs, with \tp\ measured from the optical (27 events) 
or from the GeV (3 events) light curves. 
We revise the sample of G12 with new data recently appearing in the literature, and we extend it, 
beyond GRB 110213A, including all new GRBs up to July 2016 with an {\it optical} or {\it GeV} 
afterglow light curve showing  a peak \tp. 

Bursts with a peak in their early X--ray emission are not included in our final sample 
because the X--ray can be dominated (a) by an emission component of ``internal" origin, for example, due 
to the long lasting central engine activity \citep[e.g. ][]{Ghisellini:2007wq,Genet:2007ao,Ioka:2006zm,Panaitescu:2008ix,Toma:2006rp,Nardini:2010if}
and/or (b) by bright flares \citep{Margutti:2010ef}\footnote{\cite{Liang:2010lr}, \cite{Lu:2012bs} and \cite{Wu:2011ec} include in their samples also bursts with a peak in the X--ray light curve thus 
resulting in a larger but less homogeneous and secure sample of \tp.}. 

We excluded from our final sample  
(i) bursts with a multi--peaked optical light curve at early times\footnote{In these events \G\ might 
still be estimated, but only under some assumptions on the dynamical evolution of the burst outflow  
\citep[e.g. GRB 090124 --][]{Nappo:2014lh}.}; and 
(ii) events with an optical peak preceded by a decaying light curve (e.g. GRB 100621A, GRB 080319C present in G12) 
since the early decay suggests the possible presence of a multi--peaked structure. 
The latter events, however, were included in the sample of \tpul\ (\S2.2), considering the 
earliest epoch of their optical decay. 

\subsubsection{The gold sample}

Table \ref{tabTOT} lists all the GRBs we collected. 
The ``Gold'' sample is composed of  sources with a complete set of information, namely
measured  \tp\ (col. 6) and spectral parameters (col. 3--5). 
It contains 49 events: 48 long GRBs plus the short event 090510. 
GRBs of the Gold sample have the label ``(g)'' at the end of their names reported in col. 1 of Tab. \ref{tabTOT}. 
The redshift $z$, rest frame peak energy \ep, isotropic energy and luminosity (\eiso\ and \liso, respectively) 
are given in Tab. \ref{tabTOT}. Eight out of 49 GRBs have their \tp\ measured from the GeV light curve as observed by the \fe/LAT (labelled ``L'' 
or ``SL'' for the short GRB 090510).

For GRB 990123, GRB 080319B, and GRB 090102 reported in Tab. \ref{tabTOT}, it has been proposed that the 
early optical emission (and the observed peak) is produced by either the reverse shock (RS) 
 \citep{Bloom:2009fu,Japelj:2014kb,Sari:1999bv} or by a combination of 
forward and reverse shock  \citep{Gendre:2010qf,Steele:2009qo} or even by a two-component jet structure \citep[e.g. ][ for GRB 080319B]{Racusin:2008ye}. 
We assume for these three GRBs that the peak is due to the outflow deceleration and 
include them in our sample.

\begin{figure*}
\includegraphics[width=15.8cm]{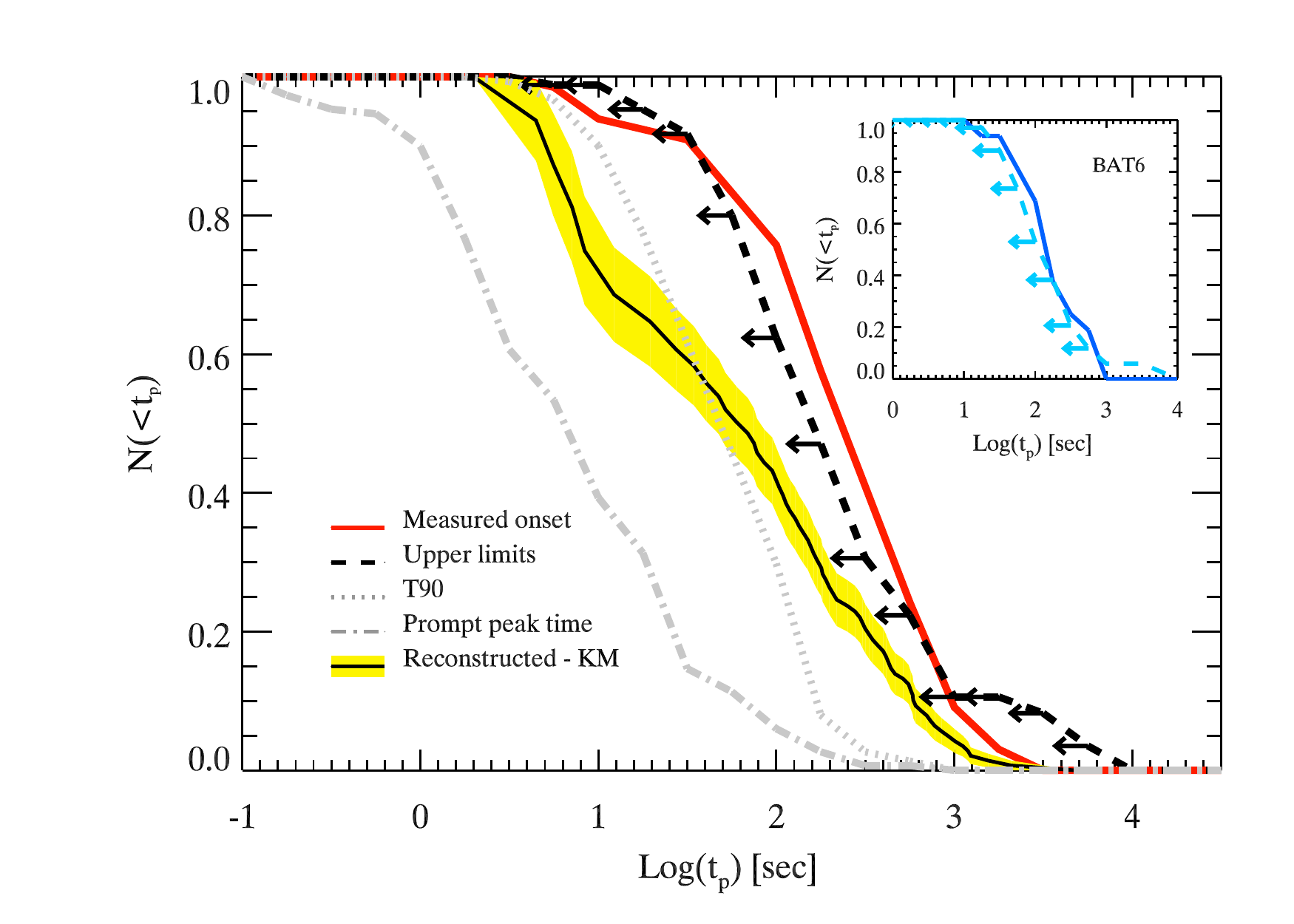}
\vskip -0.3 cm
\caption{
Cumulative distribution of the afterglow onset time \tp\ (red solid line) in the observer frame for the 
66  long GRBs of the ``Gold+Silver'' sample.  
The black dashed line (marked with leftward arrows) is the cumulative distribution of 85 
upper limits on \tp\ (filtered from Tab. \ref{tabTOT} according to \tpul$\le 5\times \max$(\tp)$\sim$11500 s). 
The black solid line (shaded yellow region) is the reconstructed \tp\ distribution (95\% confidence interval) accounting for \tpul\ through the Kaplan--Meier estimator. 
The dotted and dot--dashed lines are the cumulative distributions of \tdur\ and \tpg\ 
(i.e. the time of the peak of the prompt emission light curve), respectively, for the 151
long GRBs. The distributions have been normalised to the respective sample size. 
Insert: Distribution of \tp\ (solid blue line) and of upper limits \tpul\ (dashed cyan line with leftward arrows) of the 50 GRBs of our sample (16 with \tp\ and 34 with \tpul) also present in the \sw\ BAT complete sample \citep{Salvaterra:2012xw}. 
}
\label{fg1}
\end{figure*}

\subsubsection{The silver sample}

In our search we found 18 events with \tp\ but with poorly constrained prompt emission properties (\ep, \eiso, \liso). In most of these cases, the \sw\ Burst Alert Telescope (BAT) limited (15--150 keV) energy band coupled with a relatively low flux of the source prevent us from constraining the peak of the spectrum (\ep) 
even when it lies within the BAT energy range. 
Most of these BAT spectra were fitted by a simple power law model. 
\citealt{Sakamoto:2011tw} showed that, also in these cases, the \ep\ could be derived adopting an 
empirical correlation between the  spectral index of the power law, fitted to the BAT spectrum, and  
\ep. This empirical correlation was derived and calibrated with those bursts where BAT can measure \ep. 
Alternatively, \cite{Butler:2007os,Butler:2010il} proposed a Bayesian method to recover the value of 
\ep\ for BAT spectra fitted by the simple power law model.

We adopted the values of \ep\ and \eiso\ calculated by \cite{Butler:2007os,Butler:2010il} for 15/18 GRBs in common with their list
and the \cite{Sakamoto:2011tw} relation for the remaining 3/18 events in order to exploit the measure of 
\tp\ also for these 18 bursts. 
\citealt{Sakamoto:2011tw} and \citealt{Butler:2007os,Butler:2010il} study the time-integrated spectrum of GRBs. 
We estimated the luminosity   \liso=\eiso$(P/F)$, where $P$ and $F$ are the 
peak flux and  fluence, respectively, in the 15--150 keV energy range. 
GRBs of the ``Silver'' sample are labelled  ``(s)'' in Tab. \ref{tabTOT}. 

While we made this distinction explicit for clarity, in what follows we use the total sample of \tp\ without any 
further distinction between the Gold and Silver samples. 

\subsection{Upper limits on $t_{\rm p}$}

The afterglow onset is expected within one day for typical GRB parameters (see \S4). 
Several different observational factors, however, can prevent the measure of \tp.
It is hard to construct a sample of upper limits \tpul. \citealt{Hascoet:2014ez} included some \tp\ in their analysis but without a systematic selection criterion.
 
In this paper we collect from the literature all the GRBs with known $z$ and with an 
optical counterpart observed at least three times within one day of the trigger. 
If the light curve is decaying in time we set the upper limit \tpul\ corresponding to 
the earliest optical observation. Similar criteria apply if the long-lived afterglow emission is 
detected in the GeV energy range by the LAT. For several recent bursts, highly sampled early 
light curves are available. We excluded events with complex optical emission at early 
times and selected only those with an indication of a decaying optical flux. 

The 106 GRBs with \tpul\ are reported in Tab. \ref{tabTOT}. 
For the purposes of our analysis in the 
following we use a subsample of the 85 most constraining \tpul, that is, those with 
\tpul$\le 11500$ s which corresponds to five times the largest value of \tp\ of the 
Gold+Silver sample.


\section{Sample properties}

In this Section we present the distribution of  \tp\ (in the observer and rest frame) and study the possible correlation of the rest frame \tp\ with the observables of the prompt emission.

\subsection{Distribution of the observer frame \tp}

Figure \ref{fg1} shows the cumulative distribution (red line) of the observer frame afterglow peak time \tp\ of  long
GRBs\footnote{The short GRB 090510 is not included in the distributions. 
Its onset time \tp=0.2 s  \cite{Ghirlanda:2010pb} would place it in the lowest bin of the distribution.}. 
The distribution of upper limits \tpul\ is shown by the dashed black line (with leftward arrows). 
The distribution of measured \tp\ is consistent with that of the upper limits  at the extremes, 
that is, below 30 s and above $\sim$1000 s. In particular, the low--end of the distribution of 
\tp\ is mainly composed of bursts whose onset time is provided by the LAT data. 
Considering only GRBs with measured \tp\ (red line in Fig. \ref{fg1}), the (log) average \tp$\sim$230 s 
while upper limits (dashed black line in Fig. \ref{fg1}) have a (log) average \tp$\sim$160 s. 
The relative position of the two distributions (red and black dashed) suggests that if we considered 
only \tp\ measurements \citep[as in G12;][]{Liang:2010lr,Lu:2012bs} we would miss several 
intermediate--early onsets. This is confirmed also if we consider only the GRBs present in our sample which are part of the so called ``BAT6" sample \citep{Salvaterra:2012xw}. Indeed, this high flux cut sample of 58 \sw\ GRBs is 90\% complete in redshift. There are 16 GRBs in our sample with measured \tp\ and 34 with \tpul\ in common with the BAT6 sample (i.e. 86\% of the sample). Their \tp\ distribution (and the distribution of their \tpul) is shown in the insert of Fig. \ref{fg1}. Similarly to the larger sample, the distribution of \tp\ for the BAT6 sample is close to that of upper limits \tpul. 

In our sample nearly half of the bursts have \tp\ measured and half are upper limits. 
The distributions of \tp\ and \tpul\ overlap considerably ensuring that random censoring is present. 
Survival analysis \citep{Feigelson:1985jl} can be used to reconstruct the true distribution of \tp.  
We use the non--parametric Kaplan--Meier estimator (KM), 
as adapted by \cite{Feigelson:1985jl} to deal with upper limits. 
The KM reconstructed CDF is shown by 
the solid black line in Fig. \ref{fg1}. 
The 95\% confidence interval on this distribution (Miller 1981; Kalbfleish \& Prentice 1980) is shown by the yellow shaded region in Fig. \ref{fg1}. 
The median value of the CDF is $\langle$\tp$\rangle=60\pm20$ 
s (1$\sigma$ uncertainty). We verified that, considering a more stringent subsample of upper limits, that is, \tpul$\le 2\times \max$(\tp), similar CDF and average values are obtained.

\subsection{Distribution of the rest frame \tp}

\begin{figure}
\hskip -0.5truecm
\includegraphics[width=9.5cm]{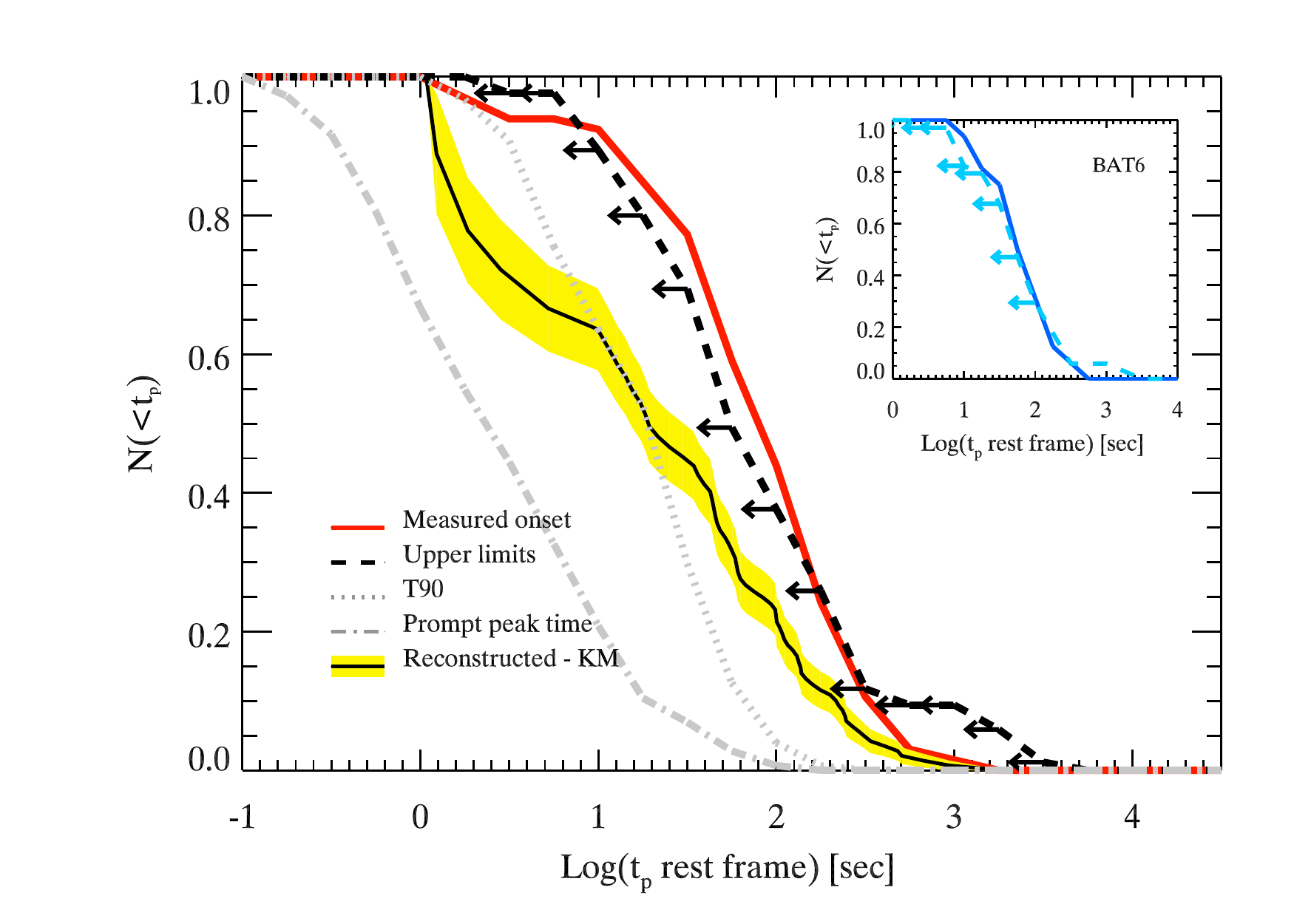}
\vskip -0.3 cm
\caption{
Cumulative distribution of the afterglow onset time \tp\ in the rest frame. Same symbols and colour code as in Fig.\ref{fg1}
}
\label{fg1bis}
\end{figure}

Figure \ref{fg1bis} shows the cumulative distribution of \tp\ in the rest frame. Colour and symbols are the same as in Fig.\ref{fg1}. In particular we note that also in the rest frame the cumulative distribution of measured \tp\ (solid red line) is close to the distribution of upper limits (leftward arrows). The KM estimator leads to a reconstructed rest frame \tp\ distribution (solid black-yellow shaded curve) which is distributed between 1 and $10^{3}$ sec with an average value of $20^{+10}_{-5}$ sec.  The insert of Fig. \ref{fg1bis} shows the distribution of the onset time of the GRBs belonging to the complete \sw\ sample. Again the measured \tp\ distribution (solid blue line) is close to the distribution of upper limits, suggesting the presence of a selection bias against the measurement of the earliest \tp\ values which is, however, not due to the requirement of the measure of the redshift.

\subsection{Comparison between \tp, $T_{90}$, and \tpg}

\begin{center}
\begin{figure}
\hskip 0.4cm
\includegraphics[width=8.8cm,height=7cm]{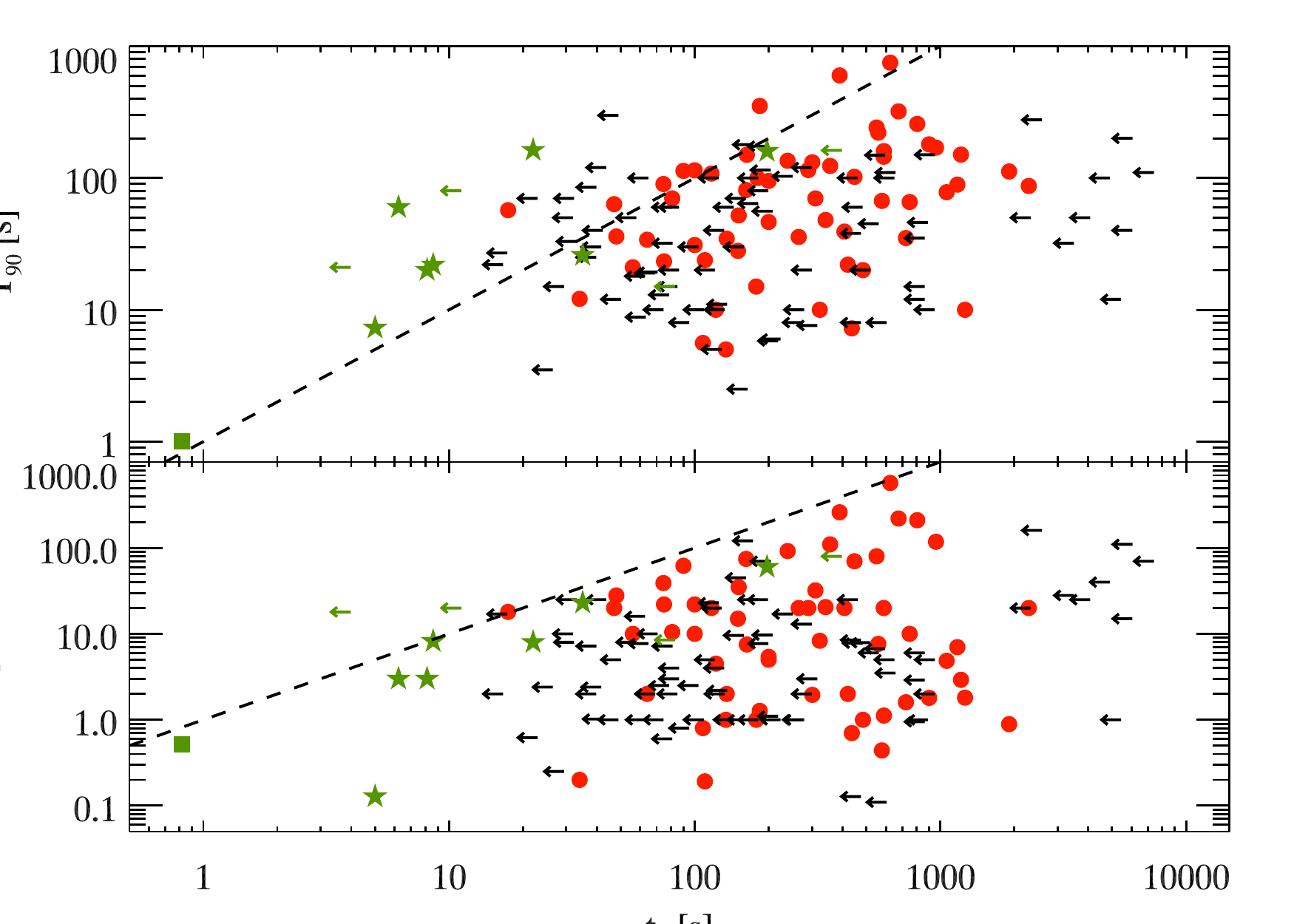}
\vskip 0.2truecm
\caption{
{\it Top panel:} GRB duration $T_{90}$ versus afterglow peak time \tp. 
GRBs with measured \tp\ are shown with red circles (green stars for  LAT bursts). 
Upper limits on \tp\ are shown by the black arrows (green arrows for LAT GRBs). 
{\it Bottom panel:} Time of the peak of the prompt emission light curve \tpg  versus \tp.  
Same symbols and colours as in the top panel. 
In both panels the equality is shown by the dashed line and the short GRB 090510 is shown by 
the green square. 
}
\label{fg2}
\end{figure}
\end{center}

One assumption for the estimate of \G\ from the measure of \tp\ (see \S4) 
is that most of the kinetic energy of the ejecta has been transferred to the blast 
wave (so called ``thin shell  approximation'' -- \citealt[][]{Hascoet:2014ez}) 
which is decelerated by the circumburst medium. 
Therefore, we should expect that  \tp\ be larger than 
the duration of the prompt emission, estimated by $T_{90}$.
To check this hypothesis, we collected $T_{90}$ 
for the bursts of our sample: its distribution is shown by the grey 
dotted line in Fig. \ref{fg1}. 
A scatter plot showing $T_{90}$ versus the observer frame \tp\ is shown in the top panel of Fig. \ref{fg2}: 
bursts with measured \tp\ are shown by the red filled circles (GRBs with \tp\ derived from the GeV--LAT 
light curve are shown by the star symbols), upper limits \tpul\ are also shown by the black  (green for LAT bursts) symbols. 
The majority (80\%) of GRBs lie below the equality line (dashed line in Fig. \ref{fg2}) having 
\tp$>T_{90}$.  20\% of the bursts have \tp$<T_{90}$. A generalised Spearman's rank correlation 
test (accounting also for upper limits -- \cite{Isobe:1986kn,Isobe:1990gb}) indicates no significant 
 correlation between $T_{90}$ and \tp\ (at $>3\sigma$ level of confidence).

The prompt emission of GRBs can be highly structured with multiple peaks separated by quiescent times. 
While $T_{90}$
is representative of the overall duration of the burst, 
another interesting timescale is the peak time of the prompt emission light curve \tpg.  
This time corresponds to the emission of a considerable fraction of  energy  during the 
prompt and it is worth comparing it with \tp. 
The distribution of \tpg\ is shown by the dot--dashed line in Fig. \ref{fg1}. 
The bottom panel of Fig. \ref{fg2} compares \tpg\ with \tp. 
Noteworthily, no GRB has 
\tp$<$\tpg. There are only two upper limits \tpul\ from the early follow up of the GeV light curve (green arrows in the bottom panel of Fig. \ref{fg2}) which lie above the 
equality line. However, the large uncertainties in these two bursts (090328 and 091003) on their GeV light curve at early times \citep{Panaitescu:2017qm} make them also compatible with having \tp$>$\tpg. 
Again, the Spearman's generalised test results in no significant correlation 
between \tp\ and \tpg. 


\subsection{Empirical correlations}

\begin{figure*}
\vskip -0.3 cm
\begin{center}
\includegraphics[trim={0 10.cm 0cm 0},clip]{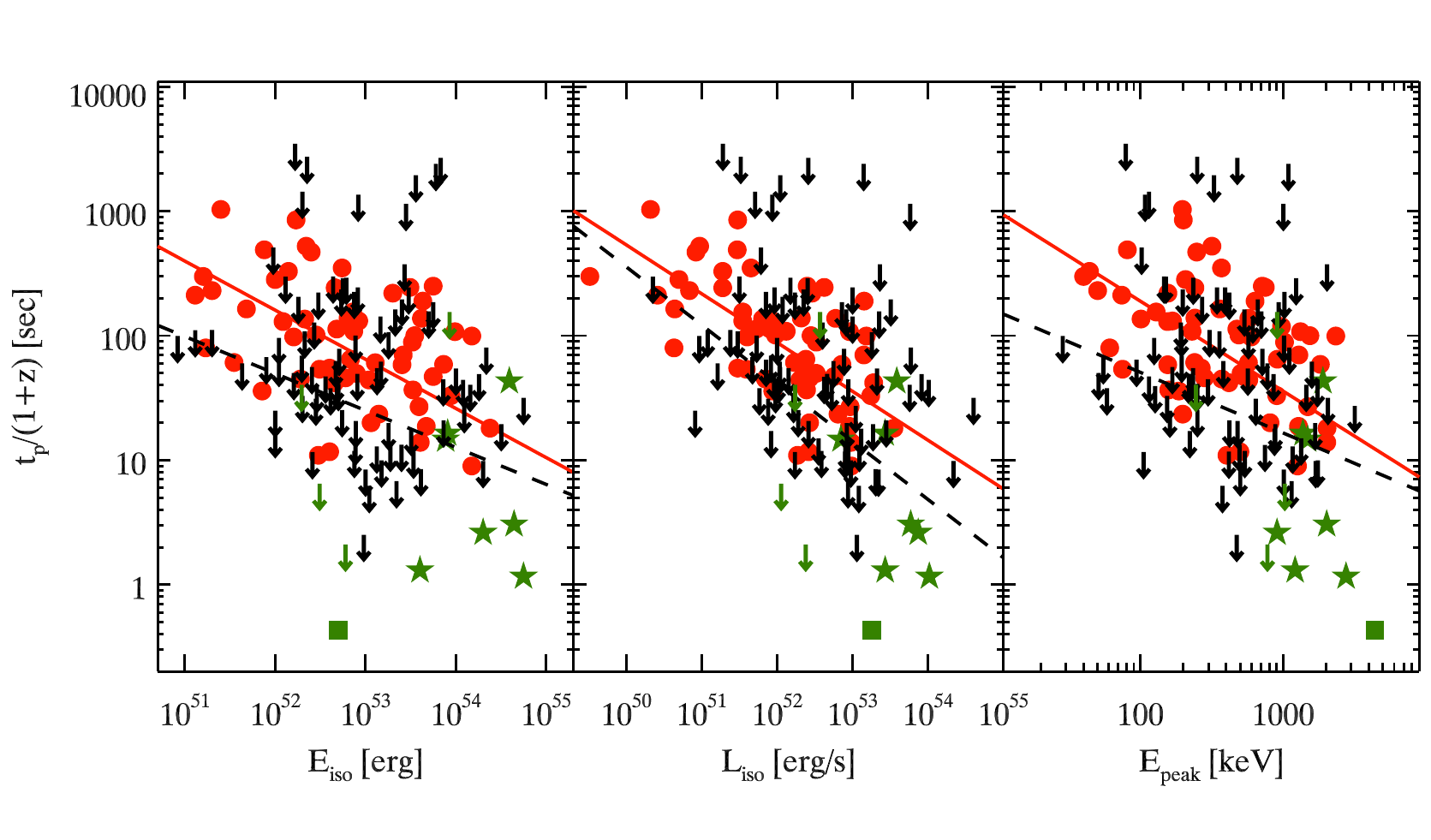}
\caption{
{\it Left panel:} \tp$/(1+z)$ versus \eiso. 
GRBs with estimates of \tp\ are shown with filled red circles (green symbols for LAT events). 
Upper limits on \tp\ (Tab. \ref{tabTOT}) are shown by the black arrows (green arrows for LAT events). 
{\it Middle panel:} \tp$/(1+z)$ versus \liso. 
{\it Right panel:}  \tp$/(1+z)$ versus \ep. 
In all panels the short GRB 090510 is shown by the green square symbol. 
}
\label{fg3}
\end{center}
\end{figure*}

We study the correlation between the onset time \tp\ in the rest frame and the energetic of GRBs. 
Figure \ref{fg3} shows \tp/$(1+z)$  versus the prompt emission isotropic energy \eiso, 
isotropic luminosity \liso\ and rest frame peak energy \ep. 

GRBs with measured \tp\ (red circles and green stars in Fig. \ref{fg3}) show significant 
correlations (chance probabilities $<10^{-5}$ -- Tab. \ref{tab4}) shown by the red 
solid lines (obtained by a least square fit with the bisector method) in the panels of Fig. \ref{fg3}. 
The correlation parameters (slope and normalisation) obtained only with \tp\  are reported in the left part of Tab. \ref{tab4}.

\begin{table*}
\centering
\begin{tabular}{@{} lccccccccc @{}} 
   \hline
Correlation & \multicolumn{4}{c}{\tp\ only} &\vline &\multicolumn{4}{c}{\tp\ \& \tpul} \\  
                               &$r$   &$P$  &$m$  &$q$ &\vline &$r$ & $P$& $m$& $q$ \\
\hline
\tp/(1+z) vs \eiso &--0.54 &3$\times10^{-6}$ &--0.71$\pm$0.06 &39.70$\pm$3.28  &\vline  &--0.20 &1$\times10^{-2}$ &--0.95$\pm$0.01 &51.61$\pm$1.00 \\
\tp/(1+z) vs \liso &--0.62 &1$\times10^{-7}$ &--0.67$\pm$0.10 &36.89$\pm$5.10  &\vline  &--0.44 &2$\times10^{-8}$ &--0.87$\pm$0.02 &46.77$\pm$1.00 \\
\tp/(1+z) vs \ep   &--0.54 &3$\times10^{-6}$ &--1.25$\pm$0.12 &5.11$\pm$0.33 &\vline  &--0.24 &3$\times10^{-3}$ &--1.25$\pm$0.03 & 4.70$\pm$0.10 \\
\hline
\end{tabular}
\vskip 0.3 cm
\caption{
Correlation between rest frame \tp/(1+z) and prompt emission properties (Fig. \ref{fg3}). 
The Spearman's correlation correlation coefficient $r$ and its chance probability $P$ are reported considering only 66 
long GRBs with estimated \tp\ or including also the 85 upper limits \tpul. 
The correlation slope $m$ and intercept $q$ of the model $Y=q+mX$ with their 1$\sigma$ uncertainties are reported. }
\label{tab4}
\end{table*}

Upper limits \tpul\ (black downward arrows in Fig. \ref{fg3})  are distributed in the same 
region of the planes occupied by \tp.
Figure \ref{fg3} shows also the lower limits on \tpll\ (grey upward arrows) derived 
assuming that \tp$>$\tpg. 

The 85 GRBs without a measured \tp\ should have their onset  \tpg$<$\tp$<$\tpul  corresponding to the vertical interval limited by the 
grey and black arrows in Fig. \ref{fg3}. 
Indeed, the reconstructed distribution of \tp\ (shown by the solid black line in 
Fig. \ref{fg1}) is bracketed by the cumulative distribution of \tpg\ on the left--hand 
side (dot--dashed grey line in Fig. \ref{fg1}) and by the distribution of 
\tpul\ on the right--hand side (dashed black line in Fig. \ref{fg1}). 

In order to evaluate the correlations of Fig. \ref{fg3} combining measured \tp\ and upper/lower 
limits we adopted a Monte Carlo approach. We assume that the KM estimator provides the distribution of \tp\ of the population of GRBs shown by the solid black line in Fig. \ref{fg1}. 
For each of the 85 GRBs with upper limits we extract randomly from the reconstructed \tp\ distribution 
a value of \tp\ (requiring that the extracted value \tp$(i)$ falls within the range  
\tpg$(i)\le$\tp$(i)\le$\tpul$(i)$ -- where $i$ runs from 1 to 85) and combine them with the 
66 GRBs with measured \tp\ to compute the correlation (using the bisector method). 
We repeat this random extraction obtaining  $10^5$  random samples and compute the average values of the Spearman's rank correlation coefficient, of its chance probability, and of the slope and normalisation of the correlations (fitted to the 10$^5$ randomly generated samples). 
In Fig. \ref{fg3}, the average correlation obtained through this Monte Carlo method is shown 
by the dot-dashed black line. The average values of the rank correlation coefficient, 
its probability, and the correlation parameters (slope and normalisation) are reported in the right section of  Tab. \ref{tab4}.

These results show that significant correlations exist between the observables (i.e. fully empirical 
at this stage). A larger energy/luminosity/peak energy corresponds to an earlier \tp. 
The distribution of upper (\tpul) and lower (\tpg) limits in the planes of Fig. \ref{fg3} 
show that these planes cannot be uniformly filled with points, further supporting the 
existence of these correlations. 

\section{On the origin of the afterglow peak time \tp}

\begin{figure*}
\center
\vskip -0.3truecm
\hskip -0.9truecm
\includegraphics[width=15cm]{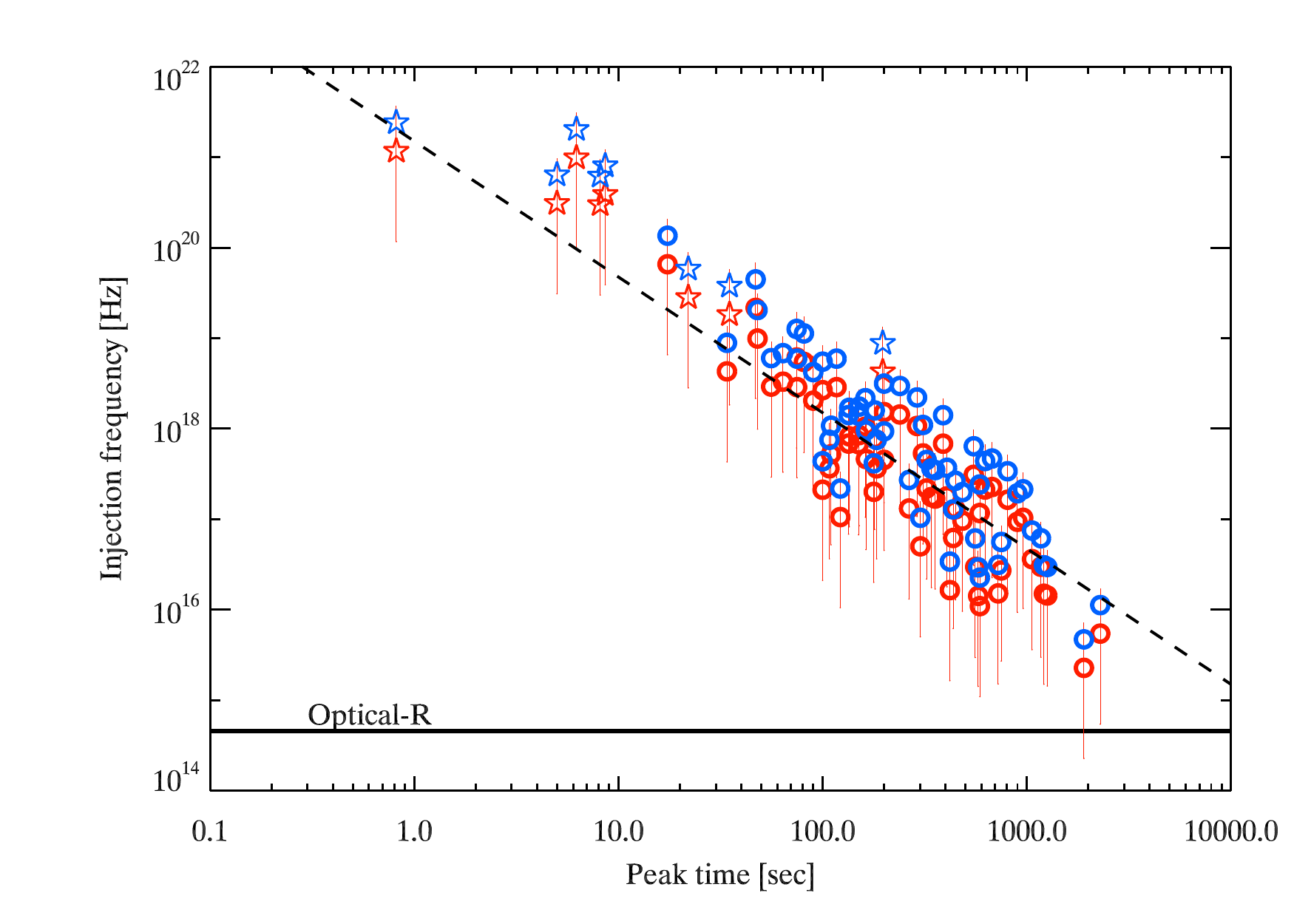}
\caption{Injection frequency at the \tp\ for each GRB. The  optical-R frequency is show by the horizontal line. The injection frequency is shown for  the homogeneous and wind case with red and blue symbols. GRBs with \tp\ from the LAT light curve are shown with star symbols. The scaling $t^{-3/2}$ of the injection frequency is shown for reference (it is not a fit) by the dashed line.  Vertical bars, shown only for the red symbols, represent the position of the injection frequency obtained assuming $\epsilon_{B}$ in the range $10^{-4}-10^{-1}$. }
\label{fg11}
\end{figure*}

In the following we assume that the afterglow peak \tp\ is produced by the fireball deceleration. However, other effects can produce an early peak in the afterglow light curve: \tp\ can be due to the passage across the observation band of the characteristic frequencies of the synchrotron spectrum. In this case, however, any of the characteristic synchrotron frequencies should lie very close to the observation band at the time of the peak. 

The synchrotron injection frequency is \citep{Panaitescu:2000fk}: 
\begin{equation}
\nu_{\rm inj}(t)=0.92\times10^{13}\epsilon_{\rm B,-2}^{1/2}\epsilon_{\rm e,-1}^{2}(E_{\rm iso,53}/\eta)^{1/2}t^{-3/2}(1+z)^{-1} \,\,\,\,\, {\rm Hz} 
\label{nuinj}
,\end{equation}
where $\epsilon_{e}$ and $\epsilon_{B}$ represent the fraction of energy shared between electrons and magnetic field at the shock and $\eta$ is the 
efficiency of conversion of kinetic energy to radiation (i.e. $E_{\rm iso, 53}/\eta$ represents the kinetic energy in units of $10^{53}$ ergs in the blast wave). Here $t$ is measured in days in the source rest frame. The above expression for $\nu_{\rm inj}(t)$ is valid either if the circumburst medium has constant density or if its density decreases with the distance from the source as $r^{-2}$ (wind medium); in the latter case only the normalisation constant is larger by a factor $\sim$2.

Figure \ref{fg11} shows $\nu_{\rm inj}(t=t_{\rm p})$ (red and blue symbols for the homogeneous or wind medium case and stars for the LAT bursts) with respect to the optical $R$ frequency (solid horizontal line). We assumed typical values of the shock parameters: $\epsilon_{B}=0.01$  and $\epsilon_{e}=0.1$ and an efficiency $\eta=$20\%  \citep{Nava:2014lr,Beniamini:2015kx}. 
The value 
$\epsilon_{\rm e}=0.1$ is consistent with \cite{Beniamini:2017lr} and \cite{Nava:2014lr} who recently found a narrow distribution of this parameter as inferred from the analysis of the radio and GeV afterglow, respectively. $\epsilon_{\rm B}$ is less constrained and has a wider dispersion, between $10^{-4}$ and $10^{-1}$  \citep{Granot:2014fk,Santana:2014qy,Zhang:2015uq,Beniamini:2016fj}, which translates into a factor of 10 for the value of $\nu_{\rm inj}$ (vertical lines in Fig. \ref{fg11})  obtained assuming $\epsilon_{\rm B}=0.01$ (open circles in Fig. \ref{fg11}) . In order to account for this uncertainty we show, as vertical lines in Fig. \ref{fg11} (for the open circles only for clarity), the possible range of frequencies that are obtained assuming $\epsilon_{\rm B}\in[10^{-4},10^{-1}]$. 
In all bursts, the injection frequency, when the afterglow peaks (i.e. at \tp), is above the optical band and it cannot produce the peak as we see it\footnote{We note also that for the LAT bursts (star symbols) the injection frequency is a factor 10 below the GeV band, and also in these cases the peak of the LAT light curve can be interpreted as the deceleration peak \citep{Nava:2017kl}. }.

However, one may argue that the above argument depends on the assumed typical values of $\epsilon_{\rm e}$ and $\epsilon_{\rm B}$. Consider, for example, the homogeneous case $s=0$. In order to ``force''  $\nu_{\rm inj}=\nu_{\rm opt,R}=4.86\times 10^{14}$ Hz at $t=t_{\rm p}$ and reproduce the observed flux at the peak $F_{\rm p}$, one would require a density of the ISM: 
\begin{equation}
n_{0}(s=0)=1.3\times10^{3}\,\, D_{28}^{4} E_{53}^{-1} \epsilon_{\rm e,-1}^{4} t_{\rm p,sec,-2}^{-3}(1+z)^{-4/3} F_{\rm p,mJy}^2\,\,\, {\rm cm^{-3}}
,\end{equation}
where $D_{28}$ is the luminosity distance in units of  $10^{28}$ cm and $t_{\rm p,sec,-2}$ is the rest frame peak time, now expressed in units of 100 seconds. 
For typical peak flux of 10 mJy we should have  densities $\sim10^{5}$ cm$^{-3}$. 

The other synchrotron characteristic frequency which could evolve and produce a peak when passing across the optical band is the cooling frequency $\nu_{\rm cool}$. For the homogeneous medium $\nu_{\rm cool}=3.7\times10^{14}\epsilon_{\rm B,-2}^{-3/2}(E_{\rm iso,53}/\eta)^{-1/2}n^{-1}t_{p}^{-1/2}(1+z)^{-1}(Y+1)^{-2}$ Hz , where $n$ is the number density of the circum burst  medium and $Y$ the Compton parameter \citep{Panaitescu:2000fk}. In this case, the portion of the synchrotron spectrum which could produce a peak is $\nu<\nu_{\rm cool}$, under the assumption that  $\nu_{\rm cool}< \nu_{\rm inj}$, and the flux evolution should be $F(t)\propto t^{1/6}$ \citep{Panaitescu:2017qm,Sari:1998lr}. This is  much shallower than the rising slopes of the afterglow emission of most of the bursts before \tp\ \citep{Liang:2013ly}. In the wind case, the cooling break increases with time $\propto t^{1/2}$ and, if transitioning accross the observing band from below, it should produce a decaying afterglow flux $F(t)\propto t^{-1/4}$ and not a peak. 

Therefore, we are confident  that \tp\ cannot be produced by any of the synchrotron frequencies crossing the optical band and it can be interpreted as due to the deceleration of the fireball and used to estimate \G. 

\section{Estimate of \G}

In this Section we revise, in chronological order,  the different methods and formulae proposed for 
estimating the bulk Lorentz factor $\Gamma_0$ \citep{Ghirlanda:2012th,Ghisellini:2010lq,Molinari:2007fj,Nava:2017kl,Sari:1999bv}. 
The scope is to compare these methods and quantify their differences.  
We  consider only the case of an adiabatic evolution of the fireball propagating in an external medium 
with a power law density profile $n(R)=n_0 R^{-s}$. The general case of a fully radiative 
(and intermediate) emission regime is discussed in \cite{Nava:2013bx}.
In \S5 we present the results, that is, estimates of \G,  for the two popular cases $s=0$ 
(homogeneous medium) and $s=2$ (wind medium). 
The estimate and comparison of \G\ in these two scenarios for a sample of 30 GRBs was presented, 
for the first time, in G12.  

During the coasting phase the bulk Lorentz factor is constant ($\Gamma=\Gamma_0$)  and the bolometric light curve 
of the afterglow scales as $L_{\rm iso} \propto t^{2-s}$. After the deceleration time, \G\ starts to decrease
and, in the adiabatic case, the light curve scales as $L_{\rm iso} \propto t^{-1}$ independently from the value of $s$.

Therefore, for a homogeneous medium ($s=0$) the light curve has a peak, while
in the wind case ($s=2$) the light curve should be flat before \tp\, and steeper afterwards\footnote{
On the other hand, as discussed in G12 and \cite{Nappo:2014lh},
there are ways to obtain a peak also in the $s=2$ case.
This is why we consider both the homogeneous and the wind case for all bursts.
Observationally, most early light curves indeed show a peak.
}. 
\G\ is the bulk Lorentz factor corresponding to the coasting phase. 
It is expected that the outflow is discontinuous with a distribution of bulk Lorentz factors 
(e.g. to develop internal shocks).  $\Gamma_0$  represents the average 
bulk Lorentz factor of the outflow during the coasting phase.

\subsection{Sari \& Piran (1999)}

In \cite[][hereafter SP99]{Sari:1999bv} $\Gamma_0$ is derived  assuming that \tp\ corresponds 
to the fireball reaching the deceleration radius $R_{\rm dec}$. 
This is  defined as the distance from the central engine, where the mass of the interstellar 
medium $m(R_{\rm dec})$, swept up by the fireball, equals  $M_0/\Gamma_0$: 
\begin{equation}
m(R_{\rm dec}) \, =\, {M_0\over \Gamma_0}\, =\, {E_0\over \Gamma_0^2 c^2 } 
,\end{equation}
where $E_0$ is the isotropic equivalent kinetic energy of the fireball after the prompt phase.

SP99 assume $t_{\rm dec} = R_{\rm dec}/(2c\Gamma_0^2)$ as the link between the deceleration 
time $t_{\rm dec}$ and $R_{\rm dec}$. 
This assumption corresponds implicitly to considering that the fireball travels up to 
$R_{\rm dec}$ with a constant bulk Lorentz factor equal to $\Gamma_0$, or, in other words, 
that the deceleration starts instantaneously at this radius. 
Instead, the deceleration of the fireball starts before $R_{\rm dec}$. 
This approximation underestimates the deceleration time $t_{\rm dec}$ and, consequently, 
underestimates $\Gamma_0$:
\begin{equation}
\Gamma_{0}^{\rm (SP99)} 
= \left[\left(\frac{3-s}{2^{5-s}\pi}\right)
\left(\frac{E_0}{n_0 m_{\rm p}c^{5-s}}\right)\right]^{\frac{1}{8-2s}}
t_{\rm p, z}^{-\frac{3-s}{8-2s}},
\label{eq:SP99}
\end{equation}
where $t_{\rm p, z}\equiv t_{\rm p}/(1+z)$. 
The original formula reported in SP99 is valid only for a homogeneous medium. 
Here, Eq. 2 has been generalised for a generic power law density profile medium.

\begin{table} 
\vskip 0.2 cm
\centering
\begin{tabular}{llcc}
  \hline
Model & $k$ &\multicolumn{2}{c}{$\qquad \Gamma_0/\Gamma_0^{\rm (N13)}$} \\
& & $s=0$ & $s=2$ \\
  \hline
SP99& $\left[\dfrac{3-s}{2^{5-s}\pi}\right]^{\frac{1}{8-2s}}$                                           &$0.852$        &$0.772$ \\
M07 & $2\left[\dfrac{3-s}{2^{5-s}\pi}\right]^{\frac{1}{8-2s}}$                                      &$1.702$         &$1.543$     \\
G10 & $\left[\dfrac{3-s}{2^{5-s}\pi(4-s)^{3-s}}\right]^{\frac{1}{8-2s}}$                                &$0.507$        &$0.649$ \\
G12 & $\left[\dfrac{17-4s}{2^{8-s}\pi(4-s)}\right]^{\frac{1}{8-2s}}$                                    &$0.687$        &$0.669$ \\
N13     & $\left[\dfrac{(17-4s)(9-2s)3^{2-s}}{2^{10-2s}\pi(4-s)}\right]^{\frac{1}{8-2s}}$       &$1$             &$1$         \\
N14 & $\left[\dfrac{(17-4s)}{16\pi(4-s)}\right]^{\frac{1}{8-2s}}$                                       &$0.971$    &$0.946$ \\
\hline
\end{tabular}
\vskip 0.3 cm
\caption{
Comparison between the different methods of estimating $\Gamma_0$ found in literature;  $k$ is a dimensionless factor that varies according to the different models and the power law index of the external medium density profile $s$ (see Eq. \ref{eq:total}). 
The values of $\Gamma_0$ are numerically compared with the one obtained by \cite{Nava:2013bx}, in the case of homogeneous medium ($s=0$) and wind medium ($s=2$). We chose to compare the result of G10 using the published formula with the parameter $a=4-s$ as prescribed by the \cite{Blandford:1976qc} dynamics. SP99: \cite{Sari:1999bv}; M07: \cite{Molinari:2007fj}; G10: \cite{Ghisellini:2010lq}; G12: \cite{Ghirlanda:2012th}; N13: \cite{Nava:2013bx}; N14: \cite{Nappo:2014lh}.
}
\label{tab:comp}
\end{table}

\subsection{Molinari et al. (2007)}

\cite[][hereafter M07]{Molinari:2007fj} introduce a new formula for $\Gamma_0$, obtained 
from SP99 by modifying the assumption on $R_{\rm dec}$. 
They, realistically, consider that the deceleration begins before $R_{\rm dec}$ so that 
$\Gamma_{\rm dec}<\Gamma_0$. Under this assumption, $m(R_{\rm dec})=M_0/\Gamma_{\rm dec}=E_0/(\Gamma_{\rm dec}^2 c^2)$ 
and $t_{\rm dec} = R_{\rm dec}/(2c\Gamma_{\rm dec}^2)$. 
However, they assume $\Gamma_{\rm dec}=\Gamma_0/2$ obtaining:
\begin{equation}\label{eq:M07}
\Gamma_{0}^{\rm (M07)} 
= 2\left[\left(\frac{3-s}{2^{5-s}\pi}\right)
\left(\frac{E_0}{n_0 m_{\rm p}c^{5-s}}\right)\right]^{\frac{1}{8-2s}}
t_{\rm p, z}^{-\frac{3-s}{8-2s}}
.\end{equation}
This estimate of \G\ is a factor 2 larger than that obtained by SP99: as discussed in \cite{Nava:2013bx}, 
M07 overestimate\footnote{\cite{Liang:2010lr}  adopt the same equation as M07.} the deceleration radius by a factor $\sim 2$ and, consequently, the value of 
$\Gamma_0$ is overestimated by the same factor. This is consistent with the results of  numerical one-dimensional (1D) simulations of the blast wave deceleration 
\citep{Fukushima:2017lr}. Their results suggest that \G\ should be a factor $\sim$2.8 smaller than that derived \citep[e.g. by][]{Liang:2010lr} through Eq. \ref{eq:M07}.

\subsection{Ghisellini et al. (2010)}

A new method to calculate the afterglow peak time \tp\ is presented in \cite[][hereafter G10]{Ghisellini:2010lq}. 
This method does not rely on the definition of the deceleration radius as in SP99 and M07. 
G10 derive \tp\ by equating the two different analytic expressions for the bolometric luminosity as a 
function of the time $L(t)$ during the coasting phase and during the deceleration phase\footnote{G10 derive  
$t_{\rm peak}$ in the case of an adiabatic or a fully radiative evolution of the fireball that propagates 
in a homogeneous medium. 
For our purposes we  consider only the adiabatic case.}.

The relation linking the radius and the time is assumed to be: $R = 2 a c t \Gamma^2$, where $a=1$ 
during the coasting phase and $a>1$ during the deceleration one. 
In the latter case, its value depends on the relation between the bulk Lorentz factor $\Gamma$ 
and the radius $R$: for an adiabatic fireball, for instance, integration of  $d R = 2c \Gamma^2 dt$, 
assuming $\Gamma \propto R^{-(3-s)/2}$ (according to the self--similar solution of \cite[][hereafter BM76]{Blandford:1976qc}, 
we obtain $a = 4-s$.
  
G10 assume a relation between $\Gamma$ and $R$ which is formally identical to the BM76 solution but with a different normalisation factor:
\begin{equation}\label{eq:gR}
\Gamma(R)=\sqrt{\frac{E_0}{m(R)c^2}}=\sqrt{\frac{(3-s)E_0}{4\pi n_0 m_p c^{2} R^{3-s}}}
,\end{equation}
where $m(R)$ is the interstellar mass swept by the fireball up to the radius $R$.

In G10, the authors are interested in the determination of the peak time of the light curve, 
but that expression can also be used to determine the initial Lorentz factor $\Gamma_0$. 
The expression, generalised for a power law profile of the external medium density, is:
\begin{equation}\label{eq:G10}
\Gamma_{0}^{\rm (G10)} 
= \left[\left(\frac{3-s}{2^{5-s}\pi(4-s)^{3-s}}\right)
\left(\frac{E_0}{n_0 m_{\rm p}c^{5-s}}\right)\right]^{\frac{1}{8-2s}}
 t_{\rm p, z}^{-\frac{3-s}{8-2s}}
.\end{equation}
This estimate of \G\ is even lower than that of SP99 and, therefore, we can reasonably presume that 
also this value of \G\ will be underestimated with respect to the real one\footnote{Although Eq. \ref{eq:G10} 
is obtained with an incorrect normalisation of the relation between $\Gamma$ and $R$, we report also 
this derivation since it was the first to propose a different method to derive the time of the peak 
of the afterglow. Cfr. with Eq. \ref{eq:BM76} showing the correct normalisation.}.

\subsection{Ghirlanda et al. (2012)}

\cite{Ghirlanda:2012th} derive another formula to estimate $\Gamma_0$, 
based on the method proposed in G10, that is, intersecting the asymptotic behaviours of 
the bolometric light curve during the coasting phase with that during the deceleration phase. 
In order to describe $\Gamma(R)$ in the deceleration regime, G12 use the BM76 solution with the 
correct (with respect to G10) normalisation factor:
\begin{equation}\label{eq:BM76}
\Gamma(R)=\sqrt{\frac{(17-4s)E_0}{(12-4s)m(R)c^2}}=\sqrt{\frac{(17-4s)E_0}{16\pi n_0 m_p c^2 R^{3-s}}}
.\end{equation}    
The relation between radius and time is that presented in G10:
\begin{equation}
 t = 
\begin{cases}
\dfrac{R}{2c \Gamma_0^2} & \mbox{if } t \ll t_{\rm p, z} \\
\\
\mbox{\LARGE$\int$}\dfrac{{\rm d}R}{2c\Gamma^2}= 
\dfrac{R}{2(4-s)c\Gamma^2} & \mbox{if } t \gg t_{\rm p, z} 
\end{cases}
,\end{equation}
where $t\ll t_{\rm p,z}$ ($t\gg t_{\rm p,z}$) corresponds to the coasting (deceleration) phase. 
The authors use these relations to obtain analytically the bolometric light curves before and 
after the peak and extrapolate them to get the intersection time which is used to infer $\Gamma_0$:
\begin{equation}\label{eq:G12}
\Gamma_{0}^{\rm (G12)} 
= \left[\left(\frac{17-4s}{2^{8-s}\pi(4-s)}\right)
\left(\frac{E_0}{n_0 m_{\rm p}c^{5-s}}\right)\right]^{\frac{1}{8-2s}}
t_{\rm p, z}^{-\frac{3-s}{8-2s}}
.\end{equation}
The main difference with respect to the formula of G10 comes from the normalisation factor 
of the BM76 solution and corresponds to a factor $[(17-4s)/(12-4s)]^{1/2}$.

\subsection{Nava et al. (2013)}

\cite[][hereafter N13]{Nava:2013bx} propose a new model to describe the dynamic evolution of the 
fireball during the afterglow emission. 
With this model, valid for an adiabatic or full and semi--radiative regime, N13 compute the 
bolometric afterglow light curves and derive 
a new analytic formula for $\Gamma_0$. They rely on the same method (intersection of 
coasting/deceleration phase luminosity solution) already used by G10 and G12, but  
with a more realistic description of the dynamics of the fireball, provide an analytic
formula for the estimate of \G\ in the case of a purely adiabatic evolution  (Eq. \ref{eq:N13}), 
and a set of numerical coefficients to be used for the full or semi--radiative evolution. 

In the rest of the present work we will adopt the formula of N13 to compute \G\ so we 
report here their equation: 
\begin{equation}\label{eq:N13}
\Gamma_{0}^{\rm (N13)} 
= \left[\frac{(17-4s)(9-2s)3^{2-s}}{2^{10-2s}\pi(4-s)}
\left(\frac{E_0}{n_0 m_{\rm p}c^{5-s}}\right)\right]^{\frac{1}{8-2s}}
t_{\rm p, z}^{-\frac{3-s}{8-2s}}
.\end{equation}
The difference with respect to the formula of G12 is due to the different relation  between the 
shock radius $R$ and the observed time $t$. 
All the preceding derivations assume that most of the observed emission comes from the single 
point of the expanding fireball that is moving exactly toward the observer, along the line of sight. 
Actually, the emitted radiation that arrives at time $t$ to the observer does not come from a single 
point, but rather from a complex surface (Equal Arrival Time Surface, EATS) that does not 
coincide with the surface of the shock front. 
N13 avoid the rather complex computation of the EATSs, adopting a relation $t(R)$ proposed by 
\cite{Waxman:1997qp} to relate radii to times. For simplicity, in the ultra--relativistic approximation:
\begin{equation}\label{eq:timeWAX}
t(R) = t_{\rm R} + t_{\rm \theta} = 
\int\left(\frac{{\rm d}R}{2\Gamma_{\rm sh}^2c}\right)\,+\frac{R}{2\Gamma^2c},
\end{equation}
where $\Gamma_{\rm sh}=2\Gamma$ is the shock Lorentz factor. 
The time is the sum of a radial time $t_{\rm R}$, that is the delay of the shock front 
with respect to the light travel time at radius $R$ and an angular time $t_{\theta}$, that is the 
delay of photons emitted at the same radius $R$ but at larger angles with respect to the line of sight. 
Using the correct dynamics, N13 obtain the relation between observed time and shock front radius:
\begin{equation}
t = 
\begin{cases}
\dfrac{3R}{4c \Gamma_0^2} & \mbox{if } t \ll t_{\rm p, z} \\
\\
\dfrac{(9-2s)R}{4(4-s)c\Gamma^2} & \mbox{if } t \gg t_{\rm p, z} 
\end{cases}
.\end{equation}
N13 use these relations to obtain analytically the bolometric light curve during the coasting and the 
deceleration phase and, from their intersection time, estimate  $\Gamma_0$ through Eq. \ref{eq:N13}.

\subsection{Nappo et al. (2014)}

\cite[][hereafter N14]{Nappo:2014lh} do not introduce directly a new formula to estimate $\Gamma_0$, 
but show a new way (valid only in the ultra--relativistic regime $\Gamma \gg 1$) to convert the 
shock radius $R$ into the observer time $t$, bypassing the problem of the calculation of the EATSs. 
They assume that most of the observed emission is produced in a ring with aperture angle 
$\sin \theta = 1/\Gamma$ around the line of sight. 
The differential form of the relation between observed time and radius can be written as
$dR = c \Gamma^2 dt$, that differs from the analogous relations of G10 and G12 by a factor $2$.

Using this simplified relation coupled to the dynamics of N13 we derive a new expression for $\Gamma_0$:
\begin{equation}\label{eq:N14}
\Gamma_{0}^{\rm (N14)} 
= \left[\left(\frac{(17-4s)}{16\pi(4-s)}\right)
\left(\frac{E_0}{n_0 m_{\rm p}c^{5-s}}\right)\right]^{\frac{1}{8-2s}}
t_{\rm p, z}^{-\frac{3-s}{8-2s}}
.\end{equation}
We show in the following paragraph that this expression provides results that are very 
similar to those obtained with the formula of N13, proving that the approximation on the 
observed times is compatible with that suggested by N13 and before by \cite{Waxman:1997qp}.

\subsection{Comparison between different methods}
 
All the previous expressions have the same dependencies on the values of $E_0$, $n_0$ and $t_{\rm p, z}$. 
They differ only by a numeric factor and can be summarised in one single expression:
\begin{equation}\label{eq:total}
\Gamma_{0}
= k \left(\frac{E_0}{n_0 m_{\rm p}c^{5-s}}\right)^{\frac{1}{8-2s}}
t_{\rm p, z}^{-\frac{3-s}{8-2s}}
,\end{equation}
where $k$ is a numeric factor that depends on the chosen method and on the power law index of the external 
medium density profile.
In Tab. \ref{tab:comp} we list the different values of $k$ for the various models and we show the 
comparison between the different estimates of $\Gamma_0$ for both a homogeneous and a wind medium. 
All the possible estimates of $\Gamma_0$ are within a factor $\sim 2$ of the estimate of N13; in 
particular the value provided by the N14 formula (Eq. \ref{eq:N14}) is similar to N13 within a few percent.

\begin{center}
\begin{figure*}
{\includegraphics[trim={1.6cm 0cm 0cm 0},clip,width=0.5\textwidth]{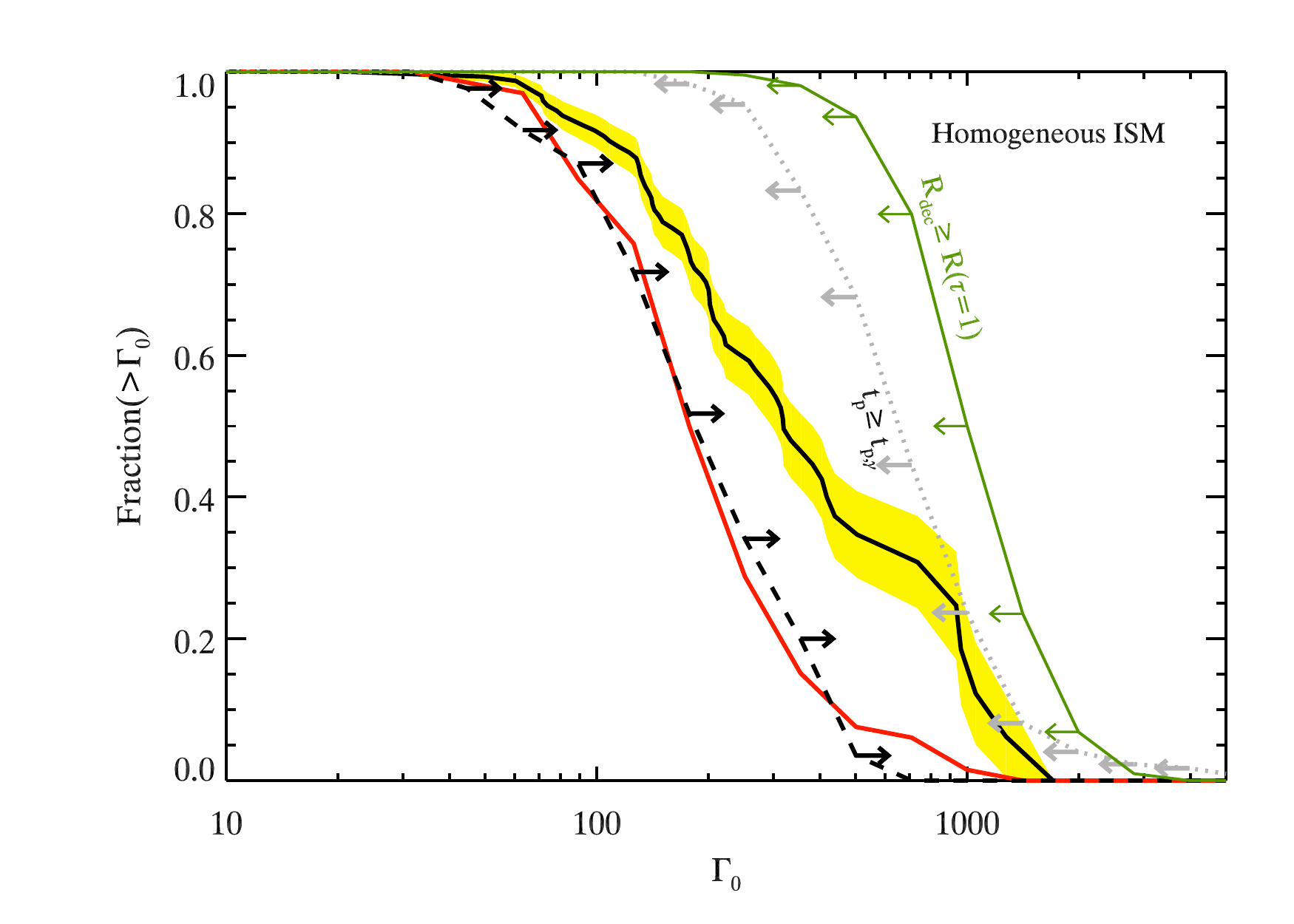}
\includegraphics[trim={2.3cm 0cm 0cm 0},clip,width=0.48\textwidth]{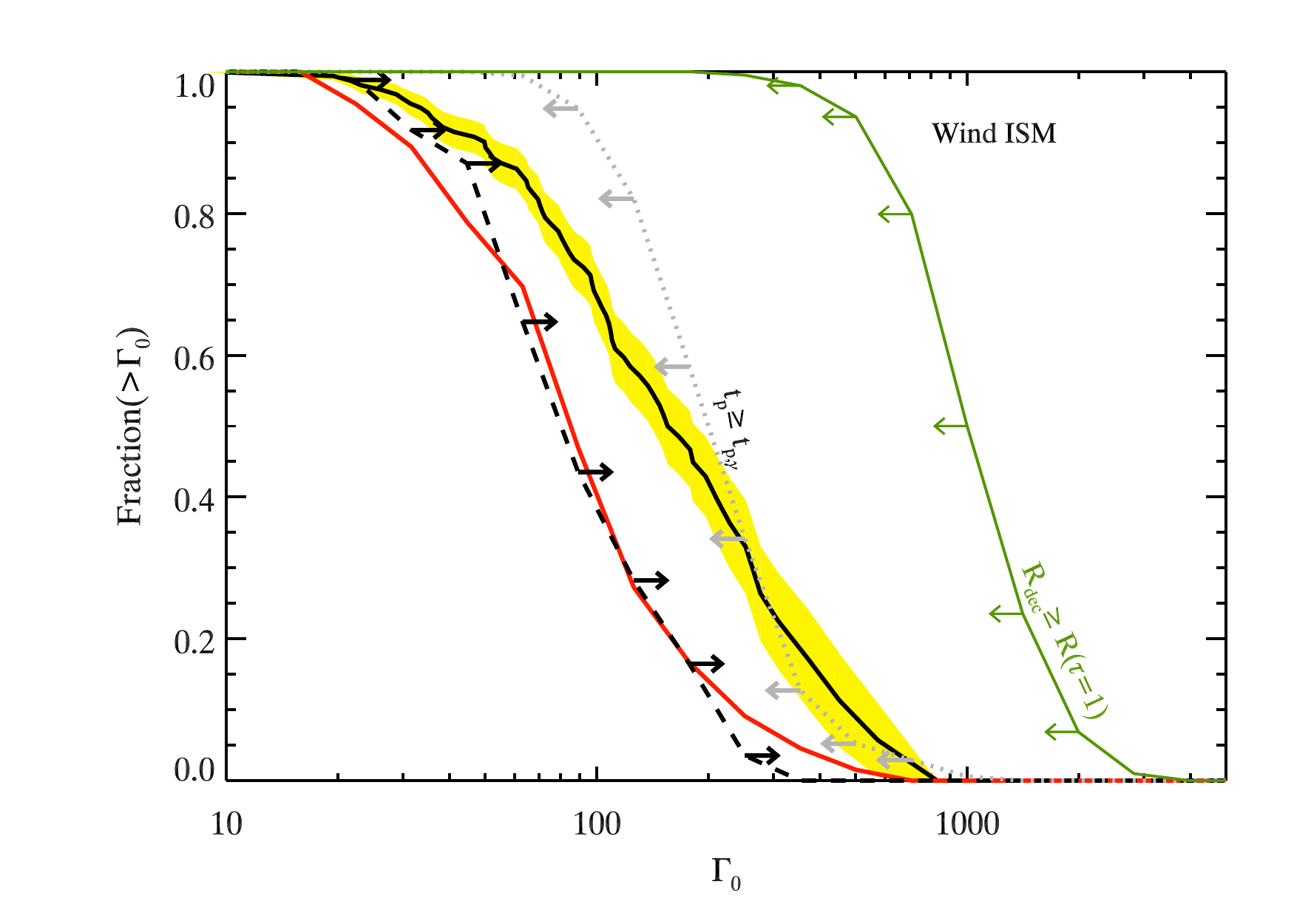}}
\caption{
Cumulative distribution of \G\ for GRBs with measured \tp\ (red solid line). 
The distribution of lower limits \GLL, derived for GRBs with an upper limit on the onset time \tpul, 
is shown by the dashed black curve (with rightward arrows). 
Assuming \tp$\ge$\tpg\ the distribution of upper limits on \G\ is shown by the dotted grey line 
(with leftward arrows). 
The most stringent limit on the distribution of \G\ is shown by the green solid line which assumes that deceleration radius 
$R_{\rm dec}\ge R(\tau=1)$, that is, the transparency radius (Eq.\ref{eqtrans}). 
Joining estimates of \G\  and lower limits \GLL\ the reconstructed (through the KM estimator) 
distribution of \G\ is shown by the solid black line (and its 95\% uncertainty by the yellow shaded region). 
The distributions are normalised to their respective number of elements. 
The tow panels (right and left, respectively) show the case of a homogeneous ($s=0$) and wind ($s=2$) medium. }
\label{fg4}
\end{figure*}
\end{center}

\section{Results}

Through Eq. \ref{eq:N13} we estimate:
\begin{itemize}
\item the bulk Lorentz factors \G\ of the 67 GRBs with measured \tp;
\item lower limits \GLL\ for the 85 bursts with upper limits on the onset time \tpul;
\item upper limit \GUL\ for the 85 bursts with lower limits on the onset time \tpll=\tpg.
\end{itemize}

We consider both a homogeneous density ISM ($s=0$) and 
a wind density profile ($s=2$).
For the first case,  we assume $n_{0}=1$ cm$^{-3}$. 
For the second case, $n(r)=n_{0}r^{-2}=\dot{M}/4\pi r^{2} m_{p} v_{w}$ 
where $\dot{M}$ is the mass-loss rate and $v_{w}$ the wind velocity. 
For typical values (e.g. Chevalier \& Li 1999)  $\dot{M}=10^{-5}M_{\odot}$ yr$^{-1}$ and $v_{w}=10^{3}$ km s$^{-1}$  
, the normalisation of the wind case is $n_{0}=10^{35} \dot{M}_{-5} v^{-1}_{w,-3}$ cm$^{-1}$.

In both cases we assume that the radiative efficiency of the prompt phase is $\eta=$20\% and 
estimate the kinetic energy of the blast wave in Eq. \ref{eq:N13} as $E_{\rm 0}=E_{\rm iso}/\eta$.   The assumed typical value for $\eta$ is similar to that reported in \cite{Nava:2014lr} and \cite{Beniamini:2015kx} who also find a small scatter of this parameter.
We notice that assuming different values of $n_{0}$ and $\eta$ within a factor of 10 and 3 with respect to those 
adopted in our analysis would introduce a systematic difference in the estimate of \G\ corresponding to a factor $\sim$1.5 (2.3) for s=0 (s=2). 

\subsection{Distribution of \G}

Fig. \ref{fg4} shows the cumulative distribution of \G. 
The solid red line is the distribution of \G\ for the 66 GRBs with a measure of \tp.  
The cumulative distribution of  \GLL\ is shown by the black dashed line 
(with rightward arrows) in Fig. \ref{fg4}. 
The cumulative distribution of \GUL\ is shown by the dotted grey 
line (with leftward arrows) in Fig. \ref{fg4}. 
We note that while \GLL\ is derived from the optical light curves decaying  
without any sign of the onset (i.e. providing \tpul), the limit \GUL\ is derived assuming that the onset 
time happens after the peak of the prompt emission (i.e. \tpg). 

A theoretical upper limit on \G\ can be derived from the the transparency radius, that is,  $R({\rm \tau=1})$
\citep{Daigne:2002lr}. The maximum bulk Lorentz factor attainable, if the acceleration is due to the internal pressure of the fireball (i.e. $R\propto \Gamma$) is:  
\begin{equation}
\Gamma_{0}\le \left(\frac{L_{\rm iso}\sigma_{T}}{8\pi m_{\rm p} c^3 \eta R_{0}}\right)^{1/4}
\label{eqtrans}
,\end{equation}
where $\sigma_{T}$ is the Thomson cross section and $R_{0}$ is the radius where the fireball is launched. 
We assume $R_{0}\sim 10^8$ cm. 
This is consistent with the value obtained from the modelling of the photospheric emission in a few GRBs 
\cite{Ghirlanda:2013fk}. 
The cumulative distribution of upper limits on \G\ obtained through Eq. \ref{eqtrans}, 
substituting for each GRB its \liso, is shown by the green solid line in Fig. \ref{fg4} 
(with the leftward green arrows). 
This distribution represents the most conservative limit on \G.\footnote{Changing 
the assumed value of $R_{0}$ within a factor of 10 shifts the green curve by a factor $\sim$1.8.}

Similarly to the cumulative distributions of \tp, shown in Fig. \ref{fg1}, also the distributions of \G\ 
(red solid line) and the distribution of lower limits \GLL\ (black dashed line) are very close to each other. 
While at low and high values of \G\ the two curves are consistent with one another, for intermediate 
values of \G\ the distribution of lower limits is very close to that of measured \G. 
In the wind case (right panel of Fig. \ref{fg4}), the lower limits distribution violates 
the distribution of measured \G. 
This suggests that the distribution of \G\ obtained only with measured \tp\ suffers from the 
observational bias related to the lack of GRBs with very early  optical observations.   

We used the KM estimator to reconstruct the distribution of \G, combining measurements and lower limits, 
similarly to what has been done in \S3.1 for \tp. 
The solid black line (with its 95\% uncertainty) in Fig. \ref{fg4} shows the most likely 
distribution of \G\ for the population of long GRBs under the assumption of a homogeneous 
ISM (left panel) and for a wind medium (right panel). 

The median values of \G\ (reported in Tab.\ref{tabG}) are 320 and 150 for the homogeneous and wind case, respectively, 
and they are consistent within their 1$\sigma$ confidence intervals. 
G12 found smaller average values of \G\ (i.e. 138 and 66 in the homogeneous and wind case, 
respectively) because of the smaller sample size (30 GRBs) and the non-inclusion of limits on \G.
Indeed, while the intermediate/small values of \G\ are reasonably well sampled by the measurements 
of \tp, the bias against  the measure of large \G\ is due to the lack of small \tp\ measurements 
(the smallest \tp\ are actually provided by the still few LAT detections). 

The reconstructed distribution of \G\ (black line in Fig. \ref{fg4}) is consistent with the distribution 
of upper limits derived assuming \tp$\ge$\tpg\  (dotted grey distribution) in the homogeneous case. 
For the wind medium there could be a fraction of GRBs ($\sim$20\%) whose \tp\ is smaller than the peak of 
the prompt emission. However, Fig. \ref{fg4} shows that, both in the homogeneous and wind case, the reconstructed \G\ 
distribution is consistent with the limiting distribution (green line) derived assuming  that 
the deceleration occurs after transparency is reached.

\begin{table}   
\centering
\begin{tabular}{lcc}
\hline
                                        &$s=0$ [68\% c.i.] &$s=2$ [68\% c.i.]   \\ 
\hline 
\G          &178 [142, 240] & 85  [65, 117] \\
\G\ \& \GLL &320 [200, 664] &155 [100, 256] \\
\hline
\end{tabular}
\caption{Average values and 68\% confidence intervals on \G\ (derived with only measured \tp\ -- red line in Fig.\ref{fg4})  
or including lower limits \GLL\ (derived from upper limits \tpul\ -- black line in Fig. \ref{fg4}).}
\label{tabG}
\end{table}

\begin{figure*}
\center
\hskip -0.1truecm
\includegraphics[width=8.5cm]{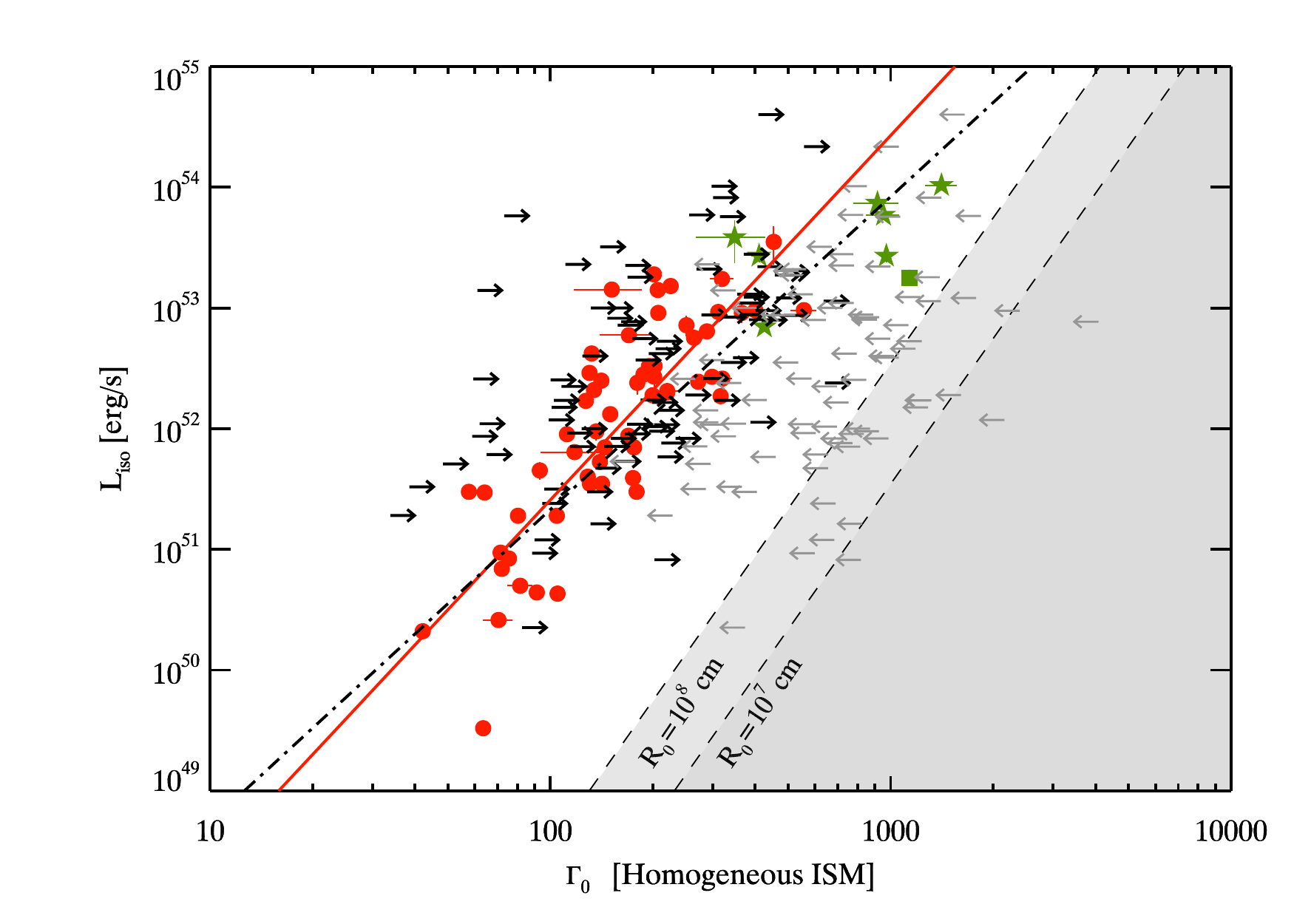}
\includegraphics[width=8.5cm]{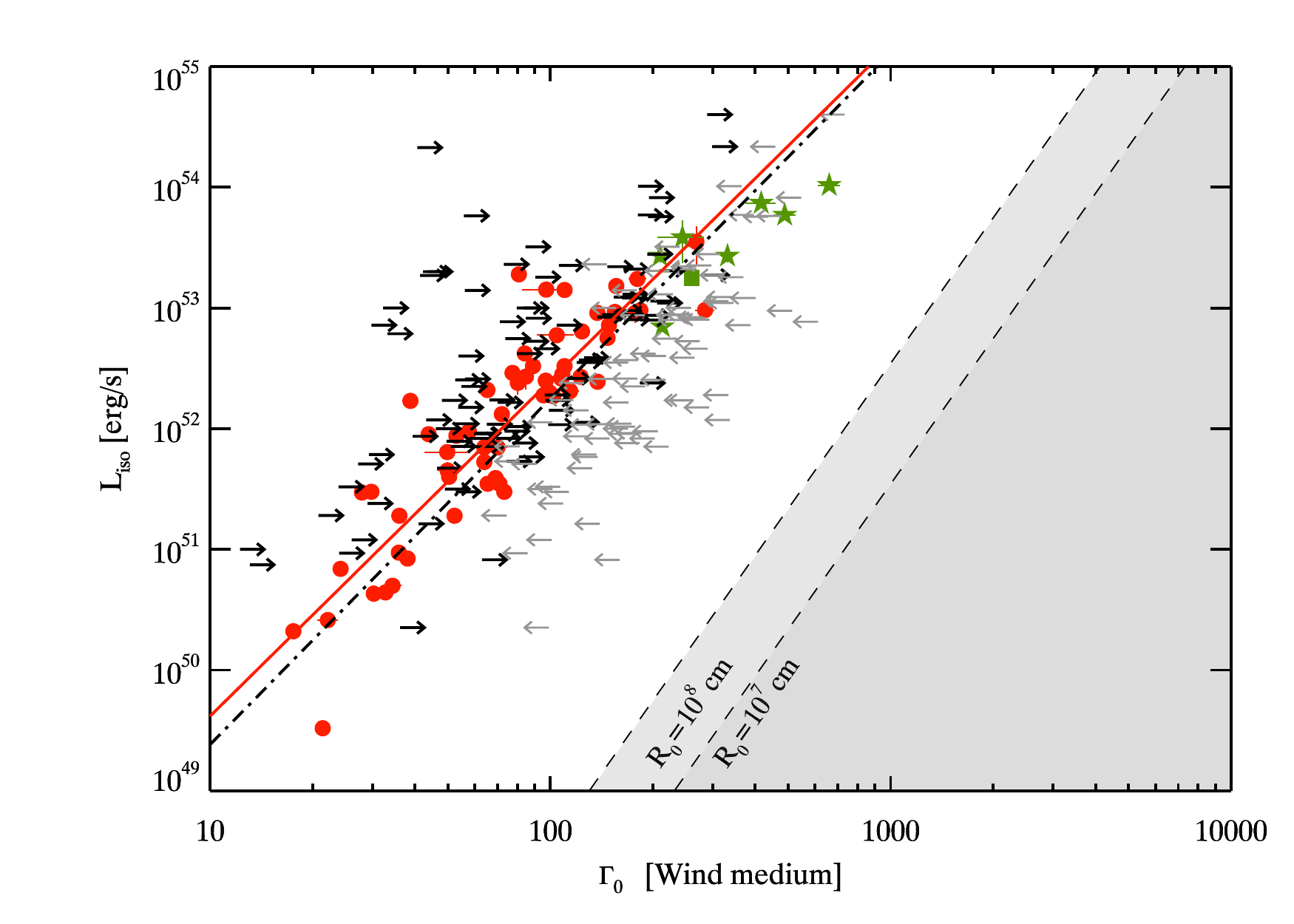}
\caption{
Correlation between the isotropic equivalent luminosity \liso\ and the bulk Lorentz factor \G. 
Estimates of \G\ from the measured afterglow onset \tp\ for 68 long GRBs (red filled circles and 
green filled stars) and the short GRB 090510 (green filled square symbol). 
Lower limits \GLL\ (derived from upper limits on the onset time \tpul) are shown by the rightward black arrows. 
Upper limits \GUL\ (derived assuming that the afterglow onset time is larger 
than the peak time of the prompt emission, that is, \tp$\ge$\tpg) are shown by the grey (leftward) arrows. 
Upper limit on \G\ imposed by the requirement that the deceleration radius is larger than the 
transparency radius is shown by the grey shaded triangular region (for two values of the radius $R_{0}$ 
where the fireball is formed -- see Eq. \ref{eqtrans}). 
The solid red line shows the correlation obtained with \G\ solely (bisector method), while the black dot--dashed line 
is the correlation obtained through the Monte Carlo method which accounts for the reconstructed distribution of \G. 
Left and right panels show the case of a homogeneous and wind medium, respectively. }
\label{fg5}
\end{figure*}
\begin{table*}
\centering
\begin{tabular}{@{} rccccccccc @{}} 
\hline\hline
\multicolumn{8}{c}{$s=0$}{$s=2$} \\
\hline
&$r$ &$P$  &$m$ &$q$ &\vline &$r$ &$P$  &$m$ &$q$ \\
\hline
\liso--\G (points) &0.86 &2$\times10^{-20}$   &3.02$\pm$0.24 &--0.59$\pm$0.10 & \vline &0.92 &1.42$\times10^{-27}$ &2.78$\pm$0.17 &--0.40$\pm$0.05 \\
(points+LL)        &0.62  &1.2$\times10^{-17}$  &2.60$\pm$0.16 &--0.67$\pm$0.06 &\vline &0.80  &1.9$\times10^{-36}$   &2.87$\pm$0.05 &--0.25$\pm$0.02 \\
\eiso--\G (points) &0.81 &3.0$\times10^{-16}$ &2.85$\pm$0.22 &--0.20$\pm$0.10 &\vline &0.93  &1.0$\times10^{-29}$  &2.66$\pm$0.13 &--1.13$\pm$0.04 \\
(points+LL)        &0.46 &2.8$\times10^{-9}$    &2.19$\pm$0.10 &--0.02$\pm$0.16 &\vline &0.73  &1.2$\times10^{-26}$   &2.66$\pm$0.05 &--0.83$\pm$0.02 \\
\ep--\G (points)   &0.73 &5$\times10^{-12}$   &1.47$\pm$0.13 &--0.24$\pm$0.06    &\vline & 0.80  & 3.3$\times10^{-16}$    &1.41$\pm$0.11 &0.25$\pm$0.04    \\
(points+LL)        &0.42 & 8$\times10^{-8}$  & 1.28$\pm$0.03 & --0.40$\pm$0.03   &\vline &0.61  &1$\times10^{-16}$     &1.43$\pm$0.03 &0.07$\pm$0.01  \\
\hline\hline
\end{tabular}
\caption{
Correlations between \liso, \eiso\ , and \ep with \G.  
Spearman's rank correlation coefficient $r$ and associated chance probability $P$, correlation 
slope $m$ and normalisation $q$ for a model $Y=mX+q$ (normalised according to Eq. \ref{norm1} and Eq. \ref{norm2}) are reported for the homogeneous
($s=0$) and wind ($s=2$) case. For each correlation the results considering only GRBs with estimated \G\ and including 
lower limits \GLL\ are given. }
\label{tabC}
\end{table*}

\section{Correlations}

G12 found correlations between the bulk Lorentz factor \G\ and the prompt emission properties of GRBs: 
\liso$\propto$\G$^{2}$, \eiso$\propto$\G$^{2}$ and, with a larger scatter, \ep$\propto$\G. 
Interestingly, combining these correlations leads to \ep$\propto$\eiso$^{0.5}$ and \ep$\propto$\liso$^{0.5}$ 
which are the \ama\  \citep{Amati:2002fy} and Yonetoku \citep{Yonetoku:2004fv} correlations. 
G12 showed that, in order to reproduce also the \ghi\ correlation \citep{Ghirlanda:2007lr},  the bulk Lorentz factor and 
the jet opening angle should be $\theta_{\rm jet}^{2}$\G=const. 

In this Section, with the 66 long GRBs with measured \G\ (a factor $\sim$3 larger sample than that used in G12)  
plus 85 lower/upper limits, we analyse the correlations of \G\ (both in the homogeneous and wind case)
 with \liso, \eiso\ and \ep. 

\begin{figure*}
\center
\hskip -0.1truecm
\includegraphics[width=8.5cm]{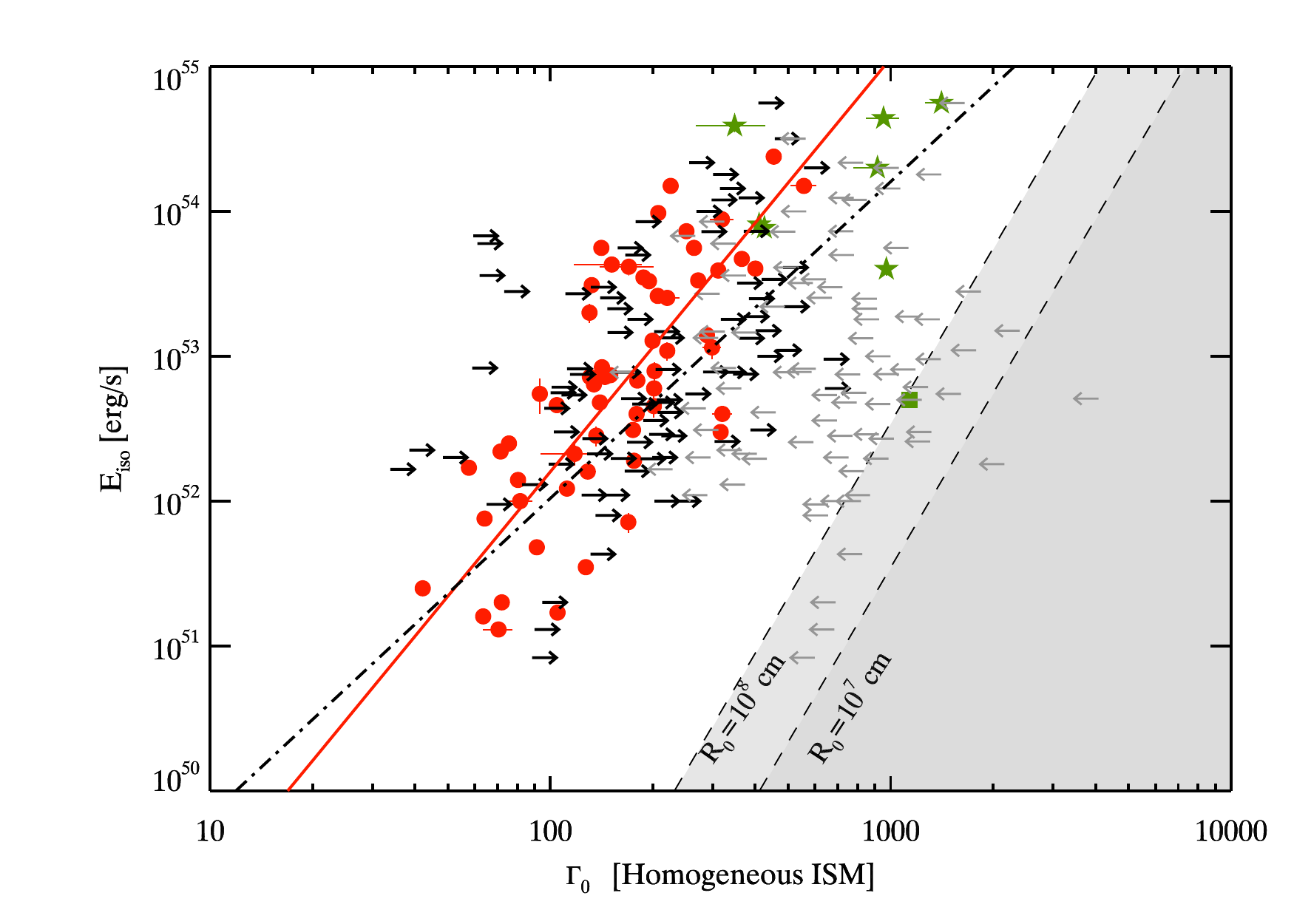}
\includegraphics[width=8.5cm]{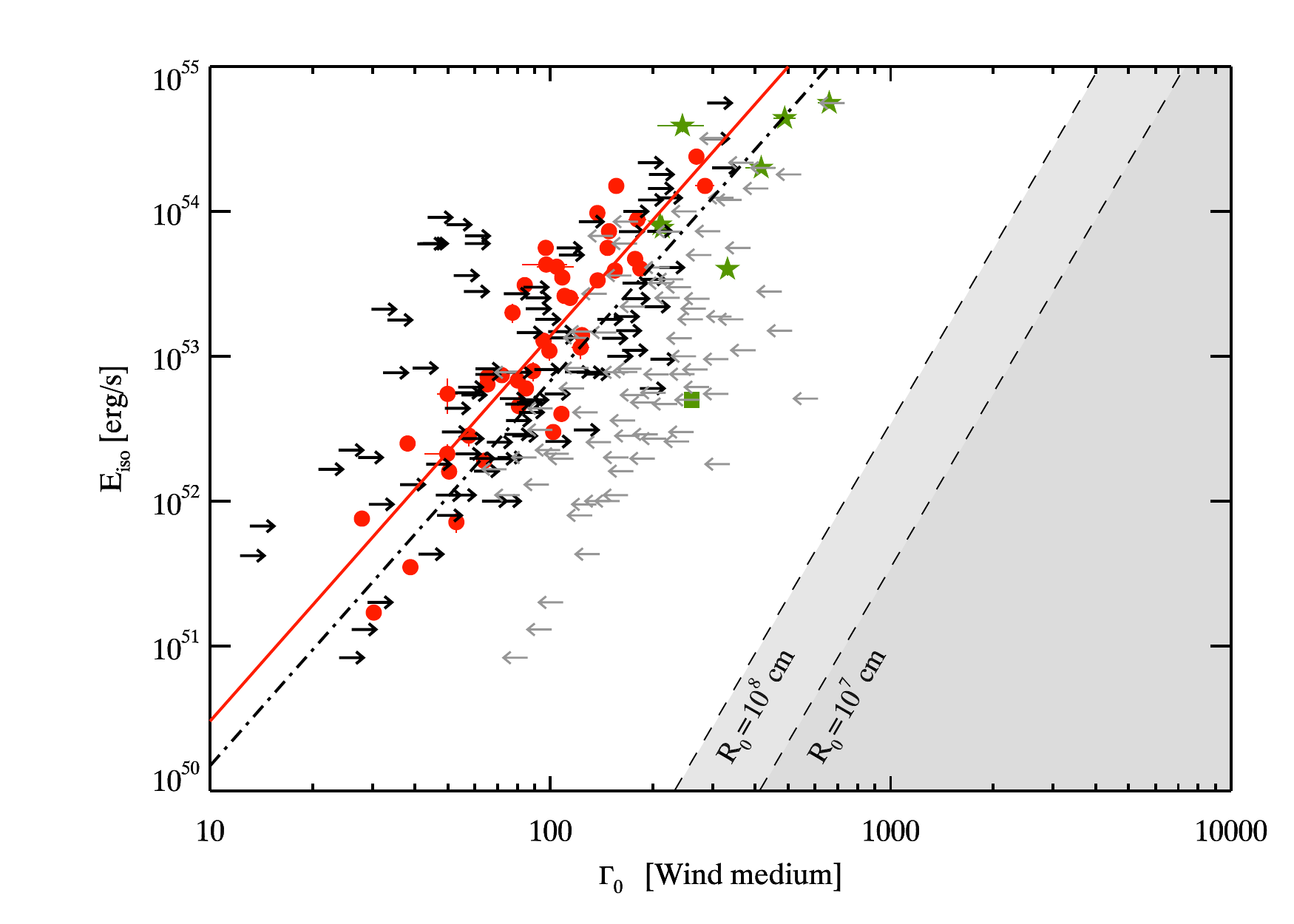}
\caption{
Correlation between the bulk Lorentz factor \G\ and the isotropic equivalent energy \eiso. Same symbols as Fig.\ref{fg5}.}
\label{fg6}
\end{figure*}
 
\begin{figure*}
\center
\hskip -0.1truecm
\includegraphics[width=8.5cm]{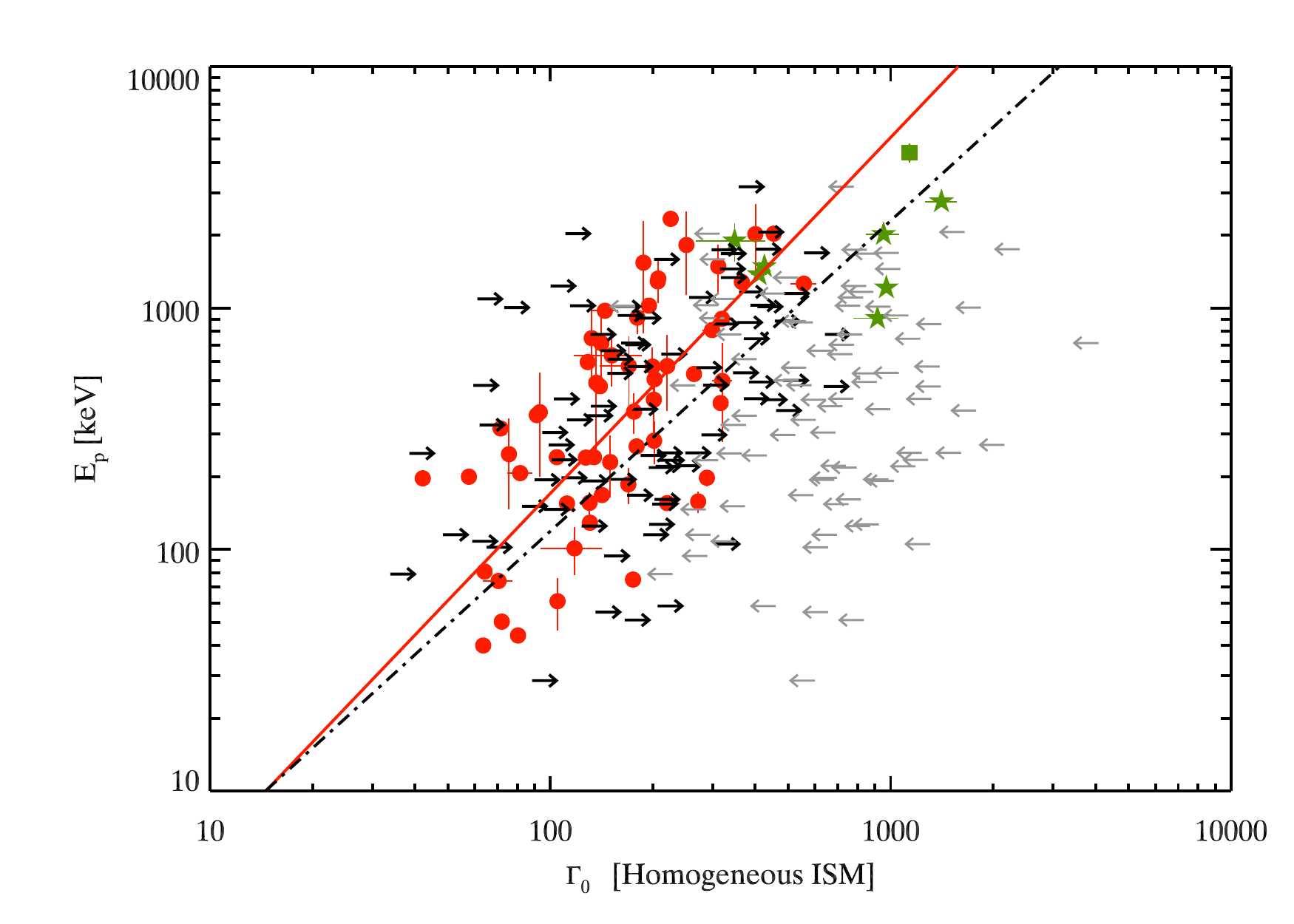}
\includegraphics[width=8.5cm]{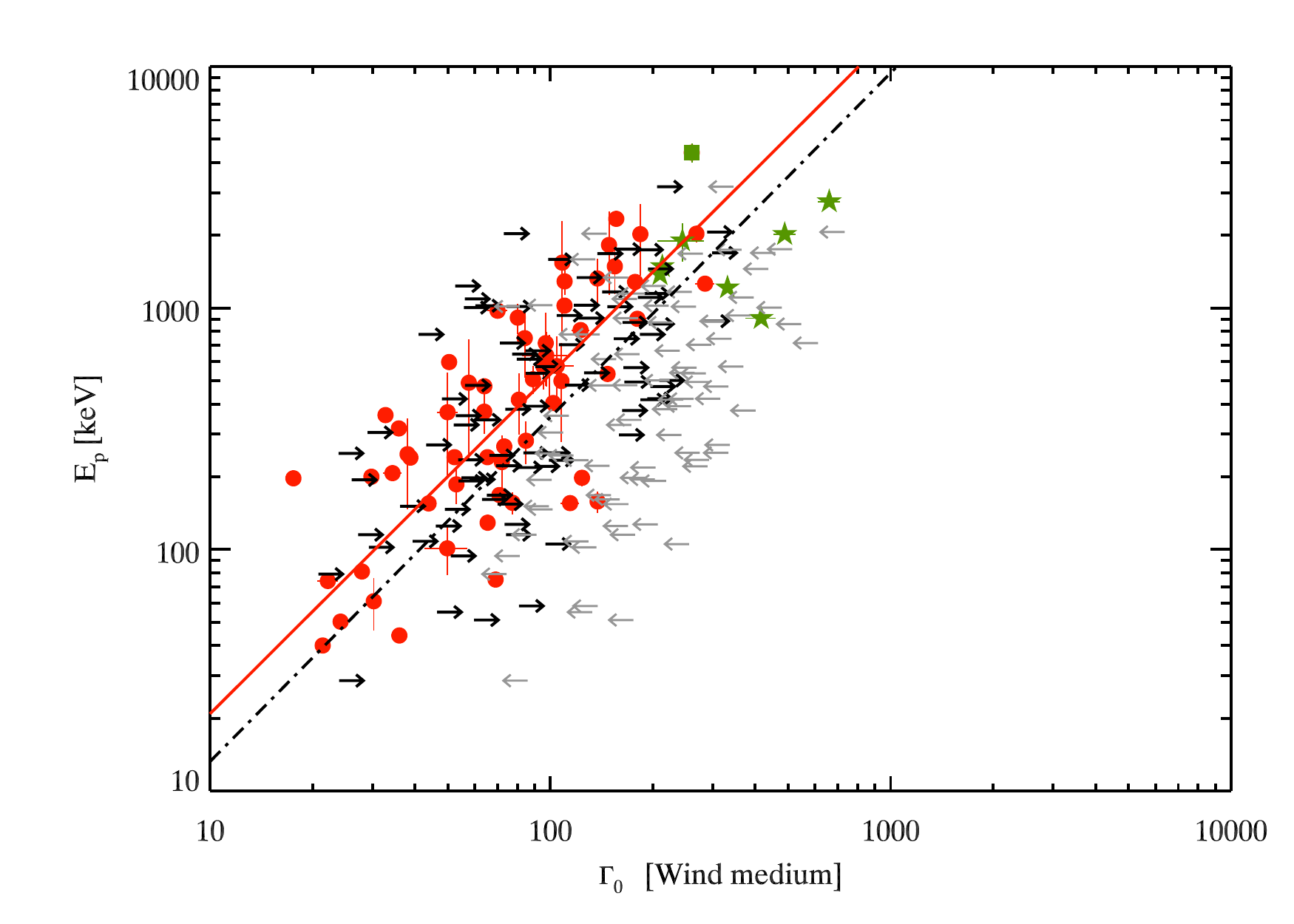}
\caption{
Correlation between the bulk Lorentz factor \G\ and the peak energy \ep. Same symbols as Fig.\ref{fg5}. }
\label{fgcom}
\end{figure*}

\begin{figure*}
\center
\hskip -0.2truecm
\includegraphics[width=8.5cm]{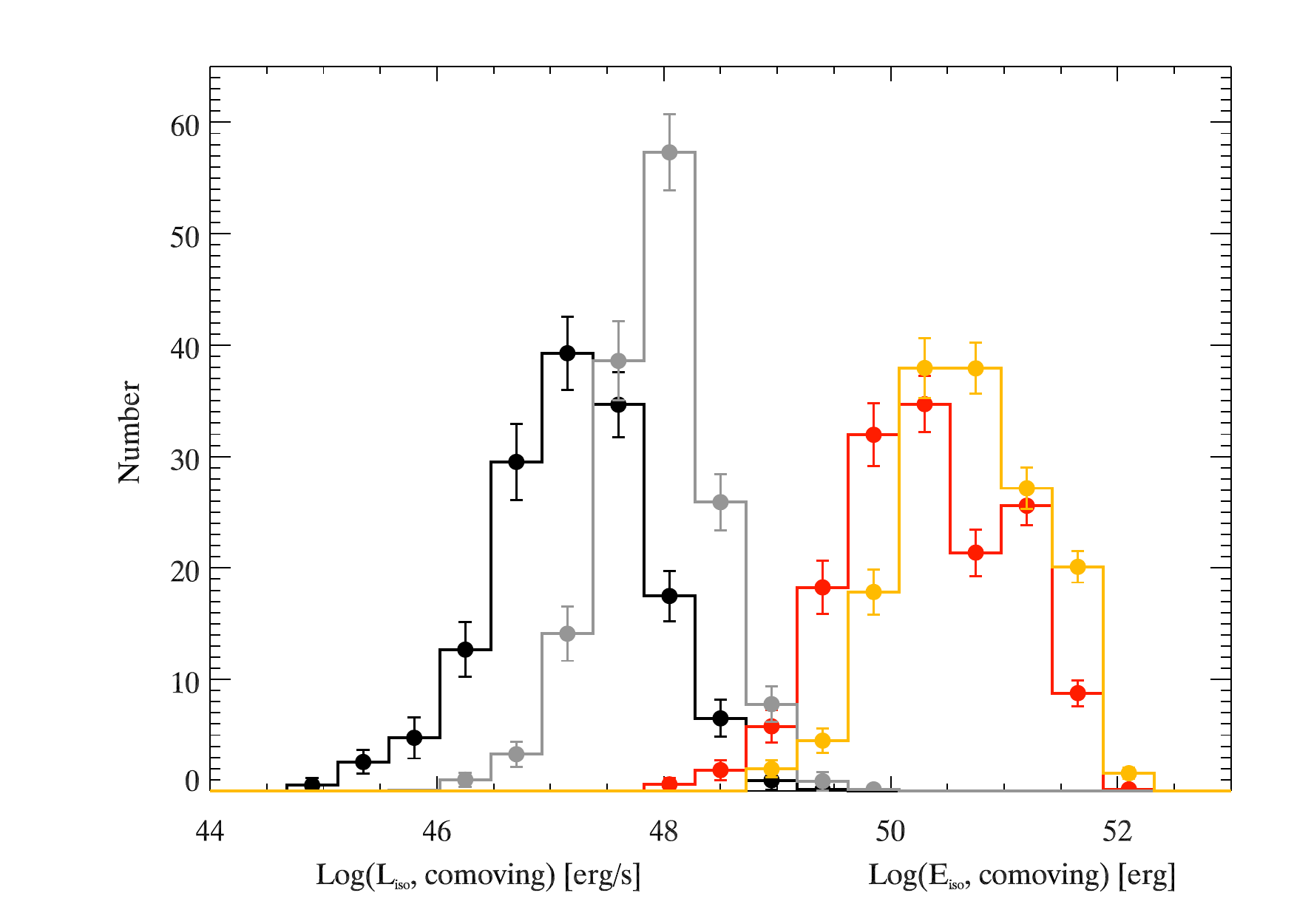}
\includegraphics[width=8.5cm]{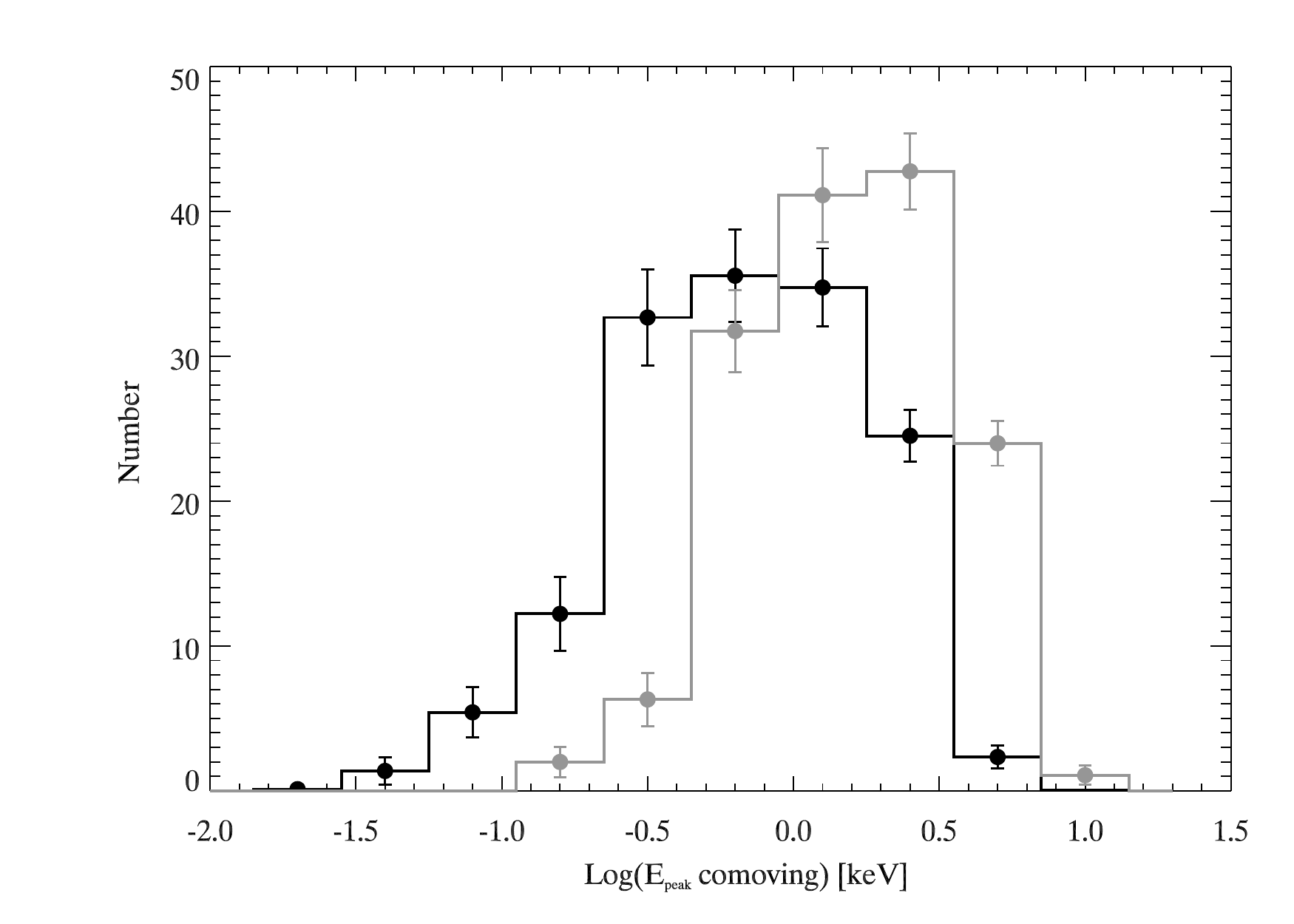}
\caption{Distribution of comoving frame properties of GRBs. Left panel: comoving frame isotropic luminosity (back and grey histogram for the homogeneous and wind case, respectively) and of the isotropic energy (red and orange histogram for the homogeneous and wind case, respectively). Right panel: comoving frame peak energy (black and grey histogram for the homogeneous and wind case, respectively). 
}\label{fg7}
\end{figure*}

Figure \ref{fg5} shows the correlation between \G\ (for the homogeneous and wind density circumburst 
medium, left and right panels, respectively) and \liso. 
Lower limits \GLL\ are shown by rightward black arrows and occupy the same region of the data points 
with estimated \G\ (red symbols). 
The green symbols show the LAT bursts which have the largest values of \liso\ and \G. 
Upper limits \GUL\ obtained requiring that the onset of the afterglow happens after 
the main emission peak of the prompt light curve are shown by the grey (leftward) arrows. 

The shadowed grey region shown in Fig. \ref{fg5} shows the limit obtained requiring that the 
deceleration happens after the transparency radius, that is, $R_{\rm dec}\ge R_{\rm th}$. 
Two different limits are shown corresponding to different (by a factor of 10) assumptions 
for $R_{0}$, that is, the radius where the jet is launched (see Eq. \ref{eqtrans}). 
The correlation between \liso\ and \G\ in the wind profile (bottom panel of Fig. \ref{fg5}) 
is less scattered than in the homogeneous case (upper panel of Fig. \ref{fg5}). 
The correlations between \G\ and the isotropic energy \eiso\  and the peak energy 
\ep\ are shown in Figs. \ref{fg6} and \ref{fgcom}, respectively. 
Symbols are the same as in Fig. \ref{fg5}. 
Also in these cases, the correlation in the wind case is less scattered than in the homogeneous case. 

The correlations shown in Figs. \ref{fg5}, \ref{fg6}, and \ref{fgcom} were analysed computing the 
Spearman's rank correlation coefficient $r$ and its probability $P$ and fitting the data, with the bisector method, 
with a linear model: 
\begin{equation}
\log\left(\frac{Y}{10^{52} {\rm erg}}\right)=m\,\log\left(\frac{\Gamma_{0}}{100}\right)+q
\label{norm1}
,\end{equation}
where $Y$ is either \liso\ or \eiso. Similarly, for the correlation between \ep\ and \G\ we adopted the model:
\begin{equation}
\log\left(\frac{E_{\rm peak}}{300 {\rm keV}}\right)=m\,\log\left(\frac{\Gamma_{0}}{100}\right)+q.
\label{norm2}
\end{equation}

The results are given in  Tab. \ref{tabC} both for the homogeneous and for the wind case. 
Correlation analysis considering only estimates of \G\ are shown by the red solid line in 
Figs. \ref{fg5}, \ref{fg6}, and \ref{fgcom}.

In order to reconstruct the correlation considering measured \G\ and upper/lower limits, 
we adopted the same Monte Carlo procedure described in \S3. 
We generate $10^5$ samples composed by the GRBs with measured \G\ and assigning to the 85 GRBs 
without measured \tp\ a value of \G\ randomly extracted from the reconstructed 
distribution of \G\ shown in Fig. \ref{fg4}. 
We then analyse the correlations of \G\ and the energetic variables within these random 
samples and report the central values of the correlation parameters (coefficient, probability, 
slope and normalisation) in Tab. \ref{tabC}. 
In all cases we find significant correlations also when upper/lower limits are accounted 
for with this Monte Carlo method.  
The corresponding correlation lines are shown with the dot--dashed lines in 
Figs. \ref{fg5}, \ref{fg6}, and \ref{fgcom}. 
We note that the correlations found with only \G\ or reconstructed accounting also 
for limits are very similar. 

Figure \ref{fg5} (and similarly Figs. \ref{fg6} and \ref{fgcom}) shows that the planes are 
not uniformly filled with data points, contrary to what is claimed by \cite{Hascoet:2014ez}. 
Indeed, the right part of the planes of Figs. \ref{fg5} and \ref{fg6} corresponding to 
large \G\ and any possible value of \liso\ and \eiso, respectively, are limited by 
the excluded region, that is, deceleration should happen after the fireball transparency. 
Moreover, the upper limits obtained requiring that the afterglow onset time \tp\ 
is after the main prompt emission peak leads to the upper limits shown by the grey 
downward arrows in Figs. \ref{fg5} and \ref{fg6}, which are even more constraining 
than the shaded regions. 
This confirms that the bulk Lorentz factor is indeed strongly correlated with the prompt 
emission properties (\eiso, \liso\ and \ep).

\begin{table*}
\centering
\begin{tabular}{@{} lcccc @{}} 
\hline
                        &\multicolumn{1}{c}{Rest Frame} & \vline  &\multicolumn{2}{c}{Comoving Frame}  \\
                        &                                                   & \vline  &\multicolumn{1}{c}{$s=0$} & \multicolumn{1}{c}{$s=2$} \\
                        &      median [68\% c.i.]                 & \vline  & median [68\% c.i.]            & median [68\% c.i.] \\
\hline
Isotropic energy [erg]   & 52.88 [52.25, 53.74]  & \vline & 50.50 [49.85, 51.33] & 50.89 [50.33, 51.57]  \\
Isotropic luminosity [erg s$^{-1}$]   & 52.40 [51.60, 53.16]  & \vline & 47.43 [46.79, 48.03] & 48.18 [47.70, 48.63] \\
Peak energy [keV]                & 2.64 [2.18, 3.08]  &  \vline & --0.008 [--0.42, 0.43]  & 0.35 [0.02, 0.69] \\
\hline
\end{tabular}
\vskip 0.2 cm
\caption{Average values and width of the distribution of the (log values of) 
\eiso, \liso\ and \ep, in the rest frame and in the comoving 
frame (both for the homogeneous and the wind density profile). }
\label{tabCOM}
\end{table*}

We compute the comoving frame peak energy and isotropic energy with the equations derived in G12:  
$E_{\rm peak}^{\prime}=3E_{\rm peak}/5\Gamma_0$ and $E_{\rm iso}^{\prime}=E_{\rm iso}/\Gamma_0$.  
The luminosity $L_{\rm iso}^{\prime}=3L_{\rm iso}/4\Gamma_{0}^{2}$.
Primed quantities refer to the comoving frame.  
For GRBs with a \tpul\  a lower limit \GLL\ transforms into an upper limit on \lisocom, \eisocom\ and \epcom. 
Vice versa, lower limits \tpll\ provide  upper limits \GUL\ which transform the rest frame 
observables into lower limits in the comoving frame. Through the Monte Carlo method adopted in the previous Sections 
we derive the distributions of the comoving frame \eisocom, \lisocom\ and \epcom; these are shown in Fig. \ref{fg7}. The average values of \G\ for the homogeneous ISM is larger than that of the wind case. For this reason the distributions of the comoving frame quantities shown in Fig. \ref{fg7} are slightly shifted in the two scenarios with the homogeneous case resulting in a slightly smaller average value of the comoving frame isotropic energy, luminosity, and peak energy. The average values of the comoving frame distributions shown in Fig. \ref{fg7} are reported in Tab. \ref{tabCOM}.

\section{Summary}

The present work assembles the largest sample of \tp, by revising and expanding (to June 2016) the original sample of 
G12, and including  upper limits \tpul\ corresponding to GRBs without a measured onset time. 
Our sample, presented in Tab. \ref{tabTOT}, is composed of:

\begin{itemize}
\item 67 GRBs with measured \tp: 66 long and one short (GRB 090510). Eight \tp\ are measured from the GeV light curve.
\item 106 long GRBs with an upper limit \tpul. 
These are GRBs detected in the optical or GeV band within one day and showing a decaying light curve. Five \tpul\ are measured from the GeV light curve. 
\end{itemize}

An upper limit \tpul\ gives a lower limit \GLL\ according to Equations  in \S4. 
Accounting only for the most stringent upper limits, we consider only the 85 bursts 
with \tpul$\lsim$11500 s, corresponding to five times the maximum \tp\ among the 66 GRBs. 
Therefore, the final sample is composed of 151 GRBs: 66 long GRBs\footnote{In most of the Figures we show for comparison also the 
short GRB 090510: this is not included in the quantitive analysis.}  
with \tp\ and 85 GRBs with \tpul. 

The observable is \tp: Figure \ref{fg1} shows that the relative position of the cumulative distribution of 
\tp\  (red line) and of  lower limits \tpul\ (dashed line) suggests the 
 presence of a selection bias against the measurement of intermediate/small \tp\ values. 
The earliest \tp\ are provided by the few GRBs with an onset measured from the GeV light curve by the LAT on board \fe.  

We can extract more information from the sample of \tp\ if we use also upper limits \tpul. 
Statistical methods \cite[e.g.][]{Feigelson:1985jl} allow us to reconstruct the distribution of an observable 
adopting measurements and upper limits, provided upper limits cover the same range of values of detections. 
This is our case as shown in Fig.\ref{fg1}. 
We adopt the Kaplan--Meier estimator to reconstruct the distribution of \tp\ of the population of long GRBs. We find that:
\begin{enumerate}

\item  the reconstructed distribution of \tp\ (solid black line with yellow shaded region in Fig. \ref{fg1}) has
median value $\langle$\tp$\rangle \sim$60 seconds extending from a few seconds (\tp\  from  LAT) to relatively late \tp$\sim 10^3$s; in the rest frame, the average \tp\/(1+z) is 20 seconds. 

\item The \tp\ distribution  is consistent with the cumulative distribution of \tpg\ (dot--dashed grey line in Fig. \ref{fg1}). \tpg\ is  
the time when the prompt emission light curve peaks. 

\item The rest frame \tp$/(1+z)$ is inversely correlated with \eiso, \liso\ and \ep\ (Fig. \ref{fg3}). 
These correlations are statistically significant (Tab. \ref{tab4}). 
\end{enumerate}

Since \G$\propto$\tp$^{-(3-s)/(8-2s)}$,  an upper limit on \tp\ provides a lower limit  \GLL. 
We combine \G\ and \GLL\ finding that: 
\begin{enumerate}[resume]

\item the reconstructed distribution of \G\ (solid black line and yellow shaded region in Fig. \ref{fg4}) has a median 
of $\langle \Gamma_{0}\rangle=$320 and 155 in the homogeneous and wind case, respectively (Tab. \ref{tabG}). 

\item \G\ values span two orders of magnitude from 20 to 1000 in the wind case (right panel in Fig. \ref{fg4}) 
and a slightly smaller range in the homogeneous case (left panel in Fig. \ref{fg4}).

\item The  distribution of \G\ is consistent with the distribution of upper limits derived under the assumption
that the afterglow peak is larger than the peak of the prompt emission (i.e. \tp$>$\tpg\ -- dotted grey line in 
Fig. \ref{fg4}).
\end{enumerate}

\begin{center}
\begin{figure}
\hskip -0.4cm
\includegraphics[width=9.5cm,height=7cm]{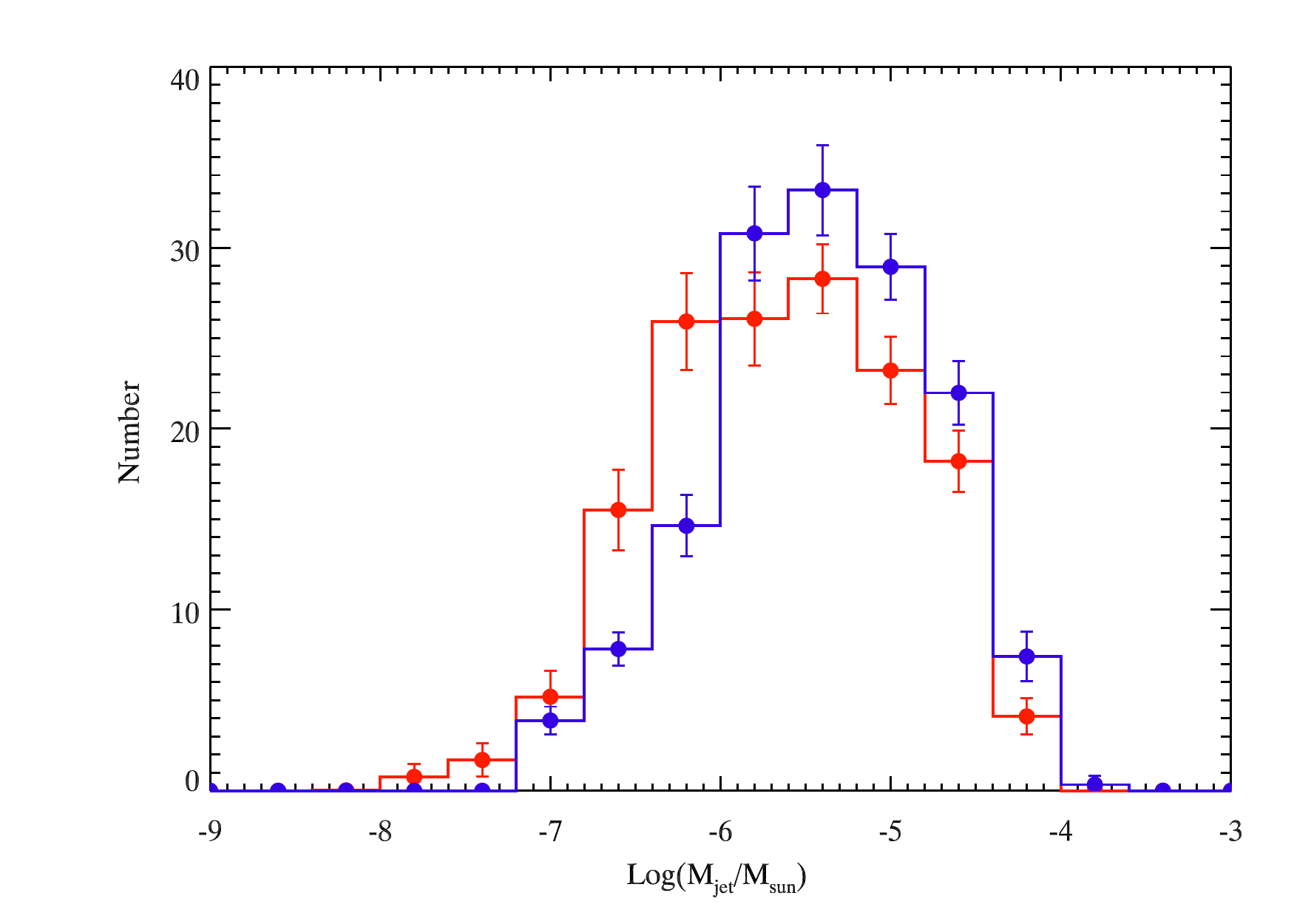}
\vskip 0.2truecm
\caption{
Distribution of the mass of the jet (in solar masses) for the homogeneous case (red solid line) and for the wind case (blue solid line) computed assuming a typical jet opening angle of 5 degrees.  
}
\label{fg9}
\end{figure}
\end{center}

Through a Monte Carlo method we combine the 66 \tp\ with the 85 GRBs with \tpul. In the latter cases 
\G\ should lie in the range \GLL$<$\G$<$\GUL\ (where \GLL\ and \GUL\ are obtained from \tpul\ and \tpg). 
We assign values of \G, randomly extracted from the reconstructed distribution and requiring they are comprised within the above limits, to the 85 GRBs. Creating $10^5$ mock samples of 66 GRBs with measured \G\ plus the randomly assigned 85 values, we evaluate the various dependencies of this parameter on the prompt emission properties analysing these random samples:  
\begin{enumerate}

\item There are significant correlations between \G\ and the prompt emission rest frame properties, 
namely \eiso, \liso\ and \ep\ (shown in Figs. \ref{fg5}, \ref{fg6}, and \ref{fg7} and Tab. \ref{tabC}):
\begin{eqnarray}
L_{\rm iso}\propto \Gamma_{0,\rm H}^{2.6} \,\,\,\,\,\,\,\,\,\,\,\, L_{\rm iso}\propto \Gamma_{0,\rm W}^{2.9} ,\\
E_{\rm iso}\propto \Gamma_{0,\rm H}^{2.2} \,\,\,\,\,\,\,\,\,\,\,\, E_{\rm iso}\propto \Gamma_{0,\rm W}^{2.7} ,\\
E_{\rm p}\propto \Gamma_{0,\rm H}^{1.3} \,\,\,\,\,\,\,\,\,\,\,\, E_{\rm p}\propto \Gamma_{0,\rm W}^{1.4} 
\label{corre}
,\end{eqnarray}
where H and W denote the homogeneous and wind case, respectively. The correlations are less scattered in the W case. 

\item The distribution of the data points and upper/lower limits in Figs. \ref{fg5}, \ref{fg6}, and \ref{fgcom} show that 
these planes are not uniformly filled with data.
\item There is no correlation between the rest frame duration $T_{90}/(1+z)$ and \G.

\end{enumerate}

With the distribution of bulk Lorentz factors reconstructed from our sample of GRBs, we can compute the baryon loading of the fireball $M=E_{\rm k}/\Gamma_{0} c^2$ where $E_{\rm k}=E_{iso}/\eta$ is the kinetic energy of the blast wave. This would lead to the isotropic equivalent mass loading but, since GRBs have a jet, we derive the baryon loading of the jet assuming a typical opening angle of 5 degrees   \citep[e.g][]{Ghirlanda:2007lr}. Figure \ref{fg9} shows the distribution of the jet baryon loading in units of solar masses obtained in the homogeneous (red line) and wind (blue line) case. The baryon loading is distributed around a typical value of a few $10^{-6} \rm{M_{\odot}}$ which is similar in the two density scenarios.  

\section{Conclusions}

We have extended the sample of GRBs with measured onset time \tp\   including for the first time also upper limits \tp. While the onset time happens relatively early (the mean value of the distribution, accounting for upper limits, is $\sim$ 60 sec in the observer frame), in most cases this happens after the time of the peak of the prompt emission of GRBs. This ensures that a considerable fraction of the fireball kinetic energy should have been transferred to the blast wave whose deceleration produces \tp. Therefore, the so called ``thin shell approximation'' (i.e. the requirement that most of the fireball energy has been transferred to the blast wave that decelerates into the interstellar medium) adopted in deriving \G\ from the measure of \tp\ holds. 
This regime is only partly established in jets that remain highly magnetised at the afterglow stage \citep{Mimica:2009fk} and seems to support a low magnetisation outflow. 
Moreover, we have shown that the synchrotron characteristic frequency $\nu_{\rm inj}$, computed at \tp, is above the optical $R$ band (Fig. \ref{fg11}) for all bursts. This ensures that it cannot be responsible for the peak of the afterglow light curve sweeping through the observer frame optical band. This allows us to interpret \tp\ as the fireball deceleration onset time and use it to compute the bulk Lorentz factor \G.  

We have reviewed and compared the different methods and formulae proposed in the literature \citep{Ghirlanda:2012th,Ghisellini:2010lq,Molinari:2007fj,Nappo:2014lh,Nava:2013bx,Sari:1999bv} to compute \G\ from \tp. They differ at most by a factor of two. Therefore, any different choice of the specific formula to compute \G\ should only introduce a small systematic difference in the derived correlation normalisation and average values.  

The average  \G\ is 320 and 155, for the homogeneous and wind cases, respectively. These values are larger than those found in G12 due to our larger sample (almost a factor 2)  and to the inclusion of lower limits \GLL. We confirm the existence of significant correlations between the GRB bulk Lorentz factor and the prompt emission observables, that is, isotropic energy and luminosity and peak energy (Eq. \ref{corre}). The \G-\liso, \G-\eiso\ and \G-\ep\ correlations are not boundaries in these 
planes (opposite to what claimed by \cite{Hascoet:2014ez}). With respect to the sample considered in \cite{Hascoet:2014ez}, our samples of \tp\  and \tpul\ are a factor three and four larger, respectively.

By combining these correlations, we find \ep$\propto$\eiso$^{0.58}$ and \ep$\propto$\liso$^{0.50}$  in the homogenous case (and \ep$\propto$\eiso$^{0.53}$ and \ep$\propto$\liso$^{0.50}$  in the wind case). These slopes are consistent with the \ama\ and \yone\ correlations \citep{Amati:2002fy,Yonetoku:2004fv} that we have re-derived here through our sample of 151 GRBs, and are $E_{\rm p}\propto E_{\rm iso}^{0.55\pm0.03}$ and $E_{\rm p}\propto L_{\rm iso}^{0.53\pm0.03}$, respectively, 

Finally, the knowledge of \G\ allows us to derive the mass of the fireball for  individual GRBs. Assuming a typical jet opening angle of 5 degrees, the jet mass $\rm M_{jet}$ is similarly distributed for the homogeneous and wind cases (red and blue lines, respectively, in Fig. \ref{fg9}) between $10^{-8}$ and $10^{-4}$ $\rm M_{\odot}$.

\newpage
\onecolumn
\begin{longtable}{@{}lllllll@{}l@{}}
\hline
\hline
GRB     &  $z$       &$\log$\ep  &$\log$\eiso   &$\log$\liso & $\log$\tp     & Ref  \\
             &               &  keV         & erg                 &erg s$^{-1}$&   s                   &       \\
\hline
  971214      &    3.42 &    2.836  $ \pm$  0.084  &    53.324  $ \pm$  0.049  &     52.858  $ \pm$  0.08   &     $\le$4.67  &  \cite{Diercks:1998qy}      &       \\ 
  980326      &    1.0  &    1.851  $ \pm$  0.220   &    51.683  $ \pm$  0.077  &     51.54   $ \pm$  0.125  &     $\le$4.56  &  \cite{Groot:1998uq}        &       \\ 
  980703      &    0.97 &    2.698  $ \pm$  0.087  &    52.839  $ \pm$  0.052  &     52.32   $ \pm$  0.101  &     $\le$4.88   &   \cite{Holland:2001fj}       &       \\ 
  990123(g)      &    1.6  &    3.308  $ \pm$  0.034  &    54.378  $ \pm$  0.051  &     53.548  $ \pm$  0.151  &     1.68  &     \cite{Galama:1999yq}         &       \\ 
  990510      &    1.62 &    2.626  $ \pm$  0.043  &    53.25   $ \pm$  0.063  &     52.787  $ \pm$  0.076  &     $\le$4.10  &     \cite{Stanek:1999kx}  &       \\ 
  990712      &    0.43 &    1.968  $ \pm$  0.070   &    51.827  $ \pm$  0.083  &     50.873  $ \pm$  0.111  &     $\le$4.18  &   \cite{Sahu:2000rt} &    \\ 
  991216      &    1.02 &    2.808  $ \pm$  0.087  &    53.829  $ \pm$  0.052  &     53.053  $ \pm$  0.144  &     $\le$4.60  &     \cite{Halpern:2000yq}      &       \\ 
  000926      &    2.07 &    2.491  $ \pm$  0.028  &    53.431  $ \pm$  0.093  &     52.675  $ \pm$  0.119  &     $\le$4.87  &    \cite{Fynbo:2001vn}     &       \\ 
  010222      &    1.48 &    2.884  $ \pm$  0.017  &    53.908  $ \pm$  0.0050 &     51.896  $ \pm$  0.025  &     $\le$4.11  &     \cite{Stanek:2001rt}     &       \\ 
  011211      &    2.14 &    2.267  $ \pm$  0.059  &    52.822  $ \pm$  0.086  &     51.501  $ \pm$  0.044  &     $\le$4.59  &     \cite{Holland:2002ys}   &       \\ 
  020124      &    3.2  &    2.591  $ \pm$  0.126  &    53.332  $ \pm$  0.147  &     52.709  $ \pm$  0.172  &     $\le$3.76  &   \cite{Berger:2002ly}  &     \\ 
  020405      &    0.69 &    2.79   $ \pm$  0.12   &    53.097  $ \pm$  0.045  &     52.14   $ \pm$  0.025  &     $\le$4.80  &     \cite{Price:2003fr}      &       \\ 
  020813      &    1.25 &    2.679  $ \pm$  0.086  &    53.831  $ \pm$  0.064  &     52.412  $ \pm$  0.04   &     $\le$3.78  &     \cite{Williams:2002ve}      &       \\ 
  021211      &    1.01 &    1.973  $ \pm$  0.088  &    52.041  $ \pm$  0.051  &     51.853  $ \pm$  0.06   &     $\le$2.11  &   \cite{Li:2003zr}  &       \\ 
  030328      &    1.52 &    2.516  $ \pm$  0.046  &    53.558  $ \pm$  0.048  &     52.041  $ \pm$  0.061  &     $\le$3.69  &   \cite{Maiorano:2006mz}      &       \\ 
  030329      &    0.17 &    1.898  $ \pm$  0.016  &    52.22   $ \pm$  0.052  &     51.281  $ \pm$  0.054  &     $\le$3.61  &     \cite{Sato:2003ly} &       \\ 
  040924      &    0.86 &    2.009  $ \pm$  0.149  &    51.978  $ \pm$  0.046  &     51.785  $ \pm$  0.078  &     $\le$2.97  &     \cite{Soderberg:2006gf} &       \\ 
  041006      &    0.72 &    2.033  $ \pm$  0.088  &    52.919  $ \pm$  0.068  &     51.937  $ \pm$  0.068  &     $\le$3.37  &    \cite{Kinugasa:2004lq}    &       \\ 
  050318      &    1.44 &    2.061  $ \pm$  0.102  &    52.301  $ \pm$  0.067  &     51.708  $ \pm$  0.068  &     $\le$3.51  &     \cite{Still:2005dq}       &       \\ 
  050401      &    2.9  &    2.7    $ \pm$  0.101  &    53.613  $ \pm$  0.085  &     53.307  $ \pm$  0.021  &     $\le$1.52  &     \cite{Melandri:2014rc}       &       \\ 
  050416A     &    0.65 &    1.456  $ \pm$  0.126  &    50.919  $ \pm$  0.152  &     50.968  $ \pm$  0.042  &     $\le$2.22  &     \cite{Melandri:2014rc}       &       \\ 
  050502A(g)     &    3.79 &    2.698  $ \pm$  0.192  &    52.602  $ \pm$  0.043  &     52.415  $ \pm$  0.05   &     1.76  &     \cite{Yost:2006rw}         &       \\ 
  050525A     &    0.61 &    2.104  $ \pm$  0.019  &    52.461  $ \pm$  0.086  &     51.979  $ \pm$  0.114  &     $\le$1.80  &     \cite{Melandri:2014rc}       &       \\ 
  050603      &    2.82 &    3.125  $ \pm$  0.035  &    53.777  $ \pm$  0.029  &     54.328  $ \pm$  0.045  &     $\le$4.52  &    \cite{Grupe:2006ve}     &       \\ 
  050730(s)      &    3.97 &    2.989  $ \pm$  0.087  &    52.857  $ \pm$  0.043  &     51.845  $ \pm$  0.043  &     2.77  &     \cite{Liang:2010lr}         &       \\ 
  050820A(g)     &    2.61 &    3.122  $ \pm$  0.091  &    53.989  $ \pm$  0.034  &     52.959  $ \pm$  0.032  &     2.59  &     \cite{Liang:2010lr}         &       \\ 
  050904      &    6.29 &    3.502  $ \pm$  0.15   &    54.093  $ \pm$  0.035  &     53.041  $ \pm$  0.154  &    $\le$2.30  &   \cite{Zaninoni:2013fk}        &       \\ 
  050908      &    3.34 &    2.29   $ \pm$  0.08   &    52.294  $ \pm$  0.071  &     51.919  $ \pm$  0.068  &     $\le$2.48  &        \cite{Zaninoni:2013fk}   &       \\ 
  050922C(g)     &    2.2  &    2.62   $ \pm$  0.123  &    52.656  $ \pm$  0.075  &     53.279  $ \pm$  0.0050 &     2.12  &    \cite{Ghisellini:2009ys}         &       \\ 
  051109A     &    2.35 &    2.732  $ \pm$  0.307  &    52.876  $ \pm$  0.051  &     52.588  $ \pm$  0.043  &     $\le$1.60  &     \cite{Yost:2007qy}     &       \\ 
  060124(g)      &    2.3  &    2.803 $ \pm$  0.111  &    53.633  $ \pm$  0.034  &     53.152  $ \pm$  0.0040 &     2.80  &     \cite{Romano:2006uq}         &       \\ 
  060206      &    4.05 &    2.581  $ \pm$  0.112  &    52.67   $ \pm$  0.066  &     52.746  $ \pm$  0.07   &     $\le$2.50  &   \cite{Melandri:2014rc}         &       \\ 
  060210(g)      &    3.91 &    2.76   $ \pm$  0.14   &    53.618  $ \pm$  0.06   &     52.775  $ \pm$  0.058  &     2.83  &     \cite{Melandri:2014rc}        &       \\ 
  060418(g)      &    1.49 &    2.757  $ \pm$  0.087  &    53.107  $ \pm$  0.034  &     52.276  $ \pm$  0.037  &     2.18  &     \cite{Molinari:2007fj}         &       \\ 
  060526      &    3.21 &    2.022  $ \pm$  0.087  &    52.412  $ \pm$  0.044  &     52.236  $ \pm$  0.078  &     $\le$1.69  &   \cite{Thone:2010ul}   &       \\ 
  060605(g)      &    3.78 &    2.69   $ \pm$  0.222  &    52.452  $ \pm$  0.069  &     51.978  $ \pm$  0.069  &     2.68  &     \cite{Rykoff:2009yq}         &       \\ 
  060607A(g)     &    3.08 &    2.76   $ \pm$  0.151  &    53.037  $ \pm$  0.062  &     52.301  $ \pm$  0.059  &     2.25  &     \cite{Molinari:2007fj}         &       \\ 
  060714      &    2.71 &    2.369  $ \pm$  0.202  &    53.127  $ \pm$  0.03   &     52.152  $ \pm$  0.031  &     $\le$2.31  &    \cite{Krimm:2007qf}      &       \\ 
  060908      &    2.43 &    2.68   $ \pm$  0.1    &    52.892  $ \pm$  0.075  &     52.415  $ \pm$  0.077  &     $\le$1.84  &     \cite{Melandri:2014rc}       &       \\ 
  060927      &    5.6  &    2.675  $ \pm$  0.107  &    52.98   $ \pm$  0.067  &     53.057  $ \pm$  0.076  &     $\le$1.22   &     \cite{Melandri:2014rc}       &       \\ 
  061007(g)      &    1.26 &    2.955  $ \pm$  0.021  &    53.945  $ \pm$  0.048  &     53.241  $ \pm$  0.061  &     1.87  &     \cite{Melandri:2014rc}        &       \\ 
  061021      &    0.35 &    2.89   $ \pm$  0.22   &    51.633  $ \pm$  0.131  &     51.212  $ \pm$  0.107  &     $\le$1.91  &     \cite{Melandri:2014rc}       &       \\ 
  061121(g)      &    1.31 &    3.11   $ \pm$  0.052  &    53.417  $ \pm$  0.05   &     53.149  $ \pm$  0.0050 &     2.21   &     \cite{Melandri:2014rc}        &       \\ 
  061126      &    1.16 &    3.126  $ \pm$  0.133  &    52.889  $ \pm$  0.045  &     52.549  $ \pm$  0.037  &     $\le$1.51   &   \cite{Perley:2008pd}      &       \\ 
  061222A     &    2.09 &    3.037  $ \pm$  0.067  &    53.778  $ \pm$  0.043  &     53.146  $ \pm$  0.117  &     $\le$3.87  &     \cite{Melandri:2014rc}       &       \\ 
  070110(g)      &    2.35 &    2.568  $ \pm$  0.2    &    52.74   $ \pm$  0.118  &     51.654  $ \pm$  0.072  &     3.07  &     \cite{Ghisellini:2009ys}   &       \\ 
  070125      &    1.55 &    2.97   $ \pm$  0.069  &    53.968  $ \pm$  0.043  &     53.511  $ \pm$  0.067  &     $\le$4.67  &     \cite{Marshall:2007fr}      &       \\ 
  070318(s)      &    0.84 &    2.556  $ \pm$  0.087  &    51.681  $ \pm$  0.043  &     50.643  $ \pm$  0.043  &     2.48  &     \cite{Liang:2010lr}       &       \\ 
  070411(s)      &    2.95 &    2.676  $ \pm$  0.087  &    52.681  $ \pm$  0.043  &     51.724  $ \pm$  0.043  &     2.65  &     \cite{Liang:2010lr}        &       \\ 
  070419A(s)     &    0.97 &    1.724  $ \pm$  0.087  &    51.204  $ \pm$  0.043  &     49.519  $ \pm$  0.043  &     2.77  &     \cite{Liang:2010lr}         &       \\ 
  071003      &    1.1  &    3.225  $ \pm$  0.06   &    53.255  $ \pm$  0.034  &     52.924  $ \pm$  0.0080 &      $\le$1.62  &   \cite{Perley:2008bh}   &       \\ 
  071010A(s)     &    0.99 &    1.869  $ \pm$  0.087  &    51.114  $ \pm$  0.043  &     50.415  $ \pm$  0.043  &     2.62  &     \cite{Covino:2008mz}         &       \\ 
  071010B(g)     &    0.95 &    2.004  $ \pm$  0.099  &    52.326  $ \pm$  0.074  &     51.806  $ \pm$  0.0040 &     2.45  &     \cite{Liang:2013ly}        &       \\ 
  071020      &    2.14 &    3.006  $ \pm$  0.088  &    53.0    $ \pm$  0.061  &     53.342  $ \pm$  0.019  &     $\le$1.40  &      \cite{Melandri:2014rc}       &       \\ 
  071025(s)      &    5.2  &    3.01   $ \pm$  0.087  &    53.519  $ \pm$  0.043  &     52.519  $ \pm$  0.043  &     2.74  &    \cite{Liang:2013ly}        &       \\ 
  071031(s)      &    2.69 &    1.643  $ \pm$  0.087  &    52.146  $ \pm$  0.043  &     51.279  $ \pm$  0.043  &     3.08  &     \cite{Liang:2013ly}         &       \\ 
  071112C*(s)    &    0.82 &    2.776  $ \pm$  0.087  &    52.204  $ \pm$  0.043  &     51.602  $ \pm$  0.043  &     2.25   &     \cite{Liang:2013ly}      &       \\ 
  080310(s)      &    2.42 &    1.875  $ \pm$  0.087  &    52.491  $ \pm$  0.043  &     51.591  $ \pm$  0.043  &     2.26  &     \cite{Liang:2013ly}         &       \\ 
  080319B(g)     &    0.94 &    3.101  $ \pm$  0.0090 &    54.176  $ \pm$  0.049  &     52.982  $ \pm$  0.01   &     1.24  &     \cite{Melandri:2014rc}         &       \\ 
  080319C     &    1.95 &    3.244  $ \pm$  0.125  &    53.176  $ \pm$  0.023  &     52.978  $ \pm$  0.0050 &     $\le$1.47  &      \cite{Melandri:2014rc}      &       \\ 
  080330(s)      &    1.51 &    1.701  $ \pm$  0.087  &    51.301  $ \pm$  0.043  &     50.839  $ \pm$  0.043  &     2.76  &      \cite{Liang:2013ly}         &       \\ 
  080413B     &    1.1  &    2.188  $ \pm$  0.094  &    52.301  $ \pm$  0.076  &     52.217  $ \pm$  0.079  &     $\le$1.84  &     \cite{Melandri:2014rc}       &       \\ 
  080603B     &    2.69 &    2.575  $ \pm$  0.088  &    53.041  $ \pm$  0.063  &     53.083  $ \pm$  0.018  &     $\le$1.36   &     \cite{Melandri:2014rc}       &       \\ 
  080605      &    1.64 &    2.823  $ \pm$  0.033  &    53.403  $ \pm$  0.062  &     53.507  $ \pm$  0.018  &     $\le$2.71  &     \cite{Melandri:2014rc}       &       \\ 
  080607      &    3.04 &    3.228  $ \pm$  0.044  &    54.301  $ \pm$  0.028  &     54.336  $ \pm$  0.2    &       $\le$1.60  &     \cite{Melandri:2014rc}       &       \\ 
  080710(s)      &    0.85 &    2.743  $ \pm$  0.087  &    51.398  $ \pm$  0.043  &     50.322  $ \pm$  0.043  &    3.28   &    \cite{Liang:2013ly}         &       \\ 
  080721      &    2.59 &    3.241  $ \pm$  0.056  &    54.079  $ \pm$  0.043  &     54.009  $ \pm$  0.064  &     $\le$2.20  &     \cite{Melandri:2014rc}       &       \\ 
  080804(g)      &    2.2  &    2.908  $ \pm$  0.024  &    53.061  $ \pm$  0.076  &     52.43   $ \pm$  0.052  &     1.80  &     \cite{Melandri:2014rc}        &       \\ 
  080810(g)      &    3.35 &    3.173  $ \pm$  0.102  &    53.592  $ \pm$  0.041  &     52.967  $ \pm$  0.041  &     2.07  &     \cite{Liang:2013ly}        &       \\ 
  080913      &    6.7  &    2.855  $ \pm$  0.261  &    52.708  $ \pm$  0.043  &     52.886  $ \pm$  0.045  &     $\le$2.78  &     \cite{Greiner:2009gf}      &       \\ 
  080916A     &    0.69 &    2.207  $ \pm$  0.105  &    52.0    $ \pm$  0.087  &     50.914  $ \pm$  0.085  &     $\le$1.62  &    \cite{Melandri:2014rc}        &       \\ 
  080916C(gL)     &    4.35 &    3.441  $ \pm$  0.019  &    54.748  $ \pm$  0.039  &     54.017  $ \pm$  0.037  &     0.79  &    \cite{Ghisellini:2010lq}         &       \\ 
  080928(s)      &    1.69 &    2.301  $ \pm$  0.087  &    52.23   $ \pm$  0.043  &     51.477  $ \pm$  0.043  &     3.36   &     \cite{Rossi:2011lr}        &       \\ 
  081007(g)      &    0.53 &    1.785  $ \pm$  0.107  &    51.23   $ \pm$  0.026  &     50.633  $ \pm$  0.04   &     2.09  &      \cite{Melandri:2014rc}        &       \\ 
  081008(s)      &    1.97 &    2.427  $ \pm$  0.087  &    52.602  $ \pm$  0.043  &     51.477  $ \pm$  0.043  &     2.21  &     \cite{Liang:2013ly}           &       \\ 
  081109A(s)     &    0.98 &    2.316  $ \pm$  0.087  &    52.0    $ \pm$  0.043  &     50.699  $ \pm$  0.043  &     2.75  &     \cite{Liang:2013ly}        &       \\ 
  081121      &    2.51 &    2.94   $ \pm$  0.061  &    53.505  $ \pm$  0.068  &     53.114  $ \pm$  0.043  &     $\le$1.80  &      \cite{Melandri:2014rc}       &       \\ 
  081203A(g)     &    2.1  &    3.188  $ \pm$  0.213  &    53.544  $ \pm$  0.037  &     52.449  $ \pm$  0.03   &     2.49  &      \cite{Melandri:2014rc}        &       \\ 
  081222      &    2.77 &    2.694  $ \pm$  0.102  &    53.398  $ \pm$  0.045  &     52.903  $ \pm$  0.067  &     $\le$1.7    &      \cite{Melandri:2014rc}       &       \\ 
  090102      &    1.55 &    3.06   $ \pm$  0.054  &    53.342  $ \pm$  0.051  &     52.94   $ \pm$  0.028  &     $\le$1.20  &      \cite{Melandri:2014rc}        &       \\ 
  090313(s)      &    3.38 &    2.382  $ \pm$  0.087  &    52.663  $ \pm$  0.043  &     51.279  $ \pm$  0.043  &     3.03  &     \cite{Melandri:2010ve}         &       \\ 
  090323(gL)      &    3.57 &    3.279  $ \pm$  0.078  &    54.591  $ \pm$  0.045  &     53.585  $ \pm$  0.167  &     2.30  &    \cite{Ghisellini:2010lq}       &       \\ 
  090328(L)      &    0.74 &    3.012  $ \pm$  0.132  &    52.491  $ \pm$  0.042  &     52.053  $ \pm$  0.012  &     $\le$1.05  &     \cite{Panaitescu:2016ul}      &       \\ 
  090418A     &    1.61 &    3.201  $ \pm$  0.165  &    53.17   $ \pm$  0.044  &     52.033  $ \pm$  0.069  &     $\le$2.21  &     \cite{Henden:2009qf}     &       \\ 
  090423      &    8.1  &    2.873  $ \pm$  0.08   &    53.274  $ \pm$  0.042  &     53.09   $ \pm$  0.029  &     $\le$2.38  &   \cite{Salvaterra:2009pd}        &       \\ 
  090424      &    0.54 &    2.225  $ \pm$  0.081  &    52.407  $ \pm$  0.051  &     52.037  $ \pm$  0.0070 &     $\le$1.94  &    \cite{Jin:2013dq}      &       \\ 
  090510(gSL)      &    0.9  &    3.643  $ \pm$  0.039  &    52.699  $ \pm$  0.043  &     53.25   $ \pm$  0.029  &     -0.09 &     \cite{Ghirlanda:2010pb}       &       \\ 
  090516      &     4.11  &    2.969  $ \pm$  0.337  &    53.748  $ \pm$  0.047  &     52.857  $ \pm$  0.016  &    $\le$3.01  &    \cite{Guidorzi:2009lq}       &       \\ 
  090618(g)      &    0.54 &    2.192  $ \pm$  0.031  &    53.403  $ \pm$  0.043  &     52.312  $ \pm$  0.018  &     1.96  &     \cite{Page:2011dq}         &       \\ 
  090709A     &    1.8  &    2.474  $ \pm$  0.039  &    53.86   $ \pm$  0.011  &     52.943  $ \pm$  0.066  &     $\le$2.10  &    \cite{Melandri:2014rc}        &       \\ 
  090715B     &    3.0  &    2.729  $ \pm$  0.133  &    53.328  $ \pm$  0.041  &     52.915  $ \pm$  0.121  &     $\le$2.80  &     \cite{Melandri:2014rc}       &       \\ 
  090812(g)      &    2.45 &    3.306  $ \pm$  0.142  &    53.605  $ \pm$  0.043  &     52.98   $ \pm$  0.044  &     1.68  &     \cite{Melandri:2014rc}        &       \\ 
  090902B(gL)     &    1.82 &    3.305  $ \pm$  0.0040 &    54.643  $ \pm$  0.0030 &     53.77   $ \pm$  0.0070 &     0.93  &       \cite{Ghisellini:2010lq}      &       \\ 
  090926A(gL)     &    2.11 &    2.958  $ \pm$  0.0030 &    54.301  $ \pm$  0.011  &     53.869  $ \pm$  0.0090 &     0.91  &     \cite{Ghisellini:2010lq}       &       \\ 
  091003(L)      &    0.9  &    2.89   $ \pm$  0.018  &    52.778  $ \pm$  0.0090 &     52.378  $ \pm$  0.011  &     $\le$0.6 &       \cite{Ghisellini:2010lq}     &       \\ 
  091018      &    0.97 &    1.74   $ \pm$  0.205  &    51.901  $ \pm$  0.051  &     51.673  $ \pm$  0.096  &     $\le$2.14  &     \cite{Melandri:2014rc}       &       \\ 
  091020(g)      &    1.71 &    2.705  $ \pm$  0.058  &    52.898  $ \pm$  0.064  &     52.519  $ \pm$  0.066  &     2.13   &     \cite{Melandri:2014rc}        &        \\ 
  091029(g)      &    2.75 &    2.362  $ \pm$  0.125  &    52.869  $ \pm$  0.043  &     52.121  $ \pm$  0.024  &     2.61  &      \cite{Filgas:2012rr}      &       \\ 
  091127      &    0.49 &    1.708  $ \pm$  0.013  &    52.207  $ \pm$  0.0    &     51.957  $ \pm$  0.011  &     $\le$1.87  &   \cite{Melandri:2014rc}         &       \\ 
  091208B(L)     &    1.06 &    2.389  $ \pm$  0.027  &    52.292  $ \pm$  0.013  &     52.238  $ \pm$  0.016  &     $\le$1.92  &   \cite{Melandri:2014rc}        &       \\ 
  100414A(gL)     &    1.39 &    3.172  $ \pm$  0.0080 &    53.886  $ \pm$  0.039  &     52.845  $ \pm$  0.043  &     1.54  &     \cite{Panaitescu:2016ul}        &       \\ 
  100621A     &    0.54 &    2.166  $ \pm$  0.068  &    52.64   $ \pm$  0.05   &     51.5    $ \pm$  0.033  &     $\le$2.65  &     \cite{Melandri:2014rc}       &       \\ 
  100728A(L)     &    1.57 &    2.958  $ \pm$  0.0080 &    53.929  $ \pm$  0.041  &     52.568  $ \pm$  0.012  &     $\le$2.599  &        \cite{Ackermann:2013fk}    &      \\ 
  100728B(g)     &    2.11 &    2.606  $ \pm$  0.031  &    52.477  $ \pm$  0.043  &     52.27   $ \pm$  0.028  &     1.53  &     \cite{Melandri:2014rc}        &       \\ 
  100814A     &    1.44 &    2.537  $ \pm$  0.019  &    52.914  $ \pm$  0.042  &     51.964  $ \pm$  0.028  &     2.77  &   \cite{Nardini:2014cr}     &    \\ 
  100901A*(b)    &    1.41 &    2.501  $ \pm$  0.087  &    52.342  $ \pm$  0.043  &     50.973  $ \pm$  0.043  &     3.1    &      \cite{Liang:2013ly}       &       \\ 
  100906A(g)     &    1.73 &    2.199  $ \pm$  0.044  &    53.524  $ \pm$  0.039  &     52.389  $ \pm$  0.015  &     2.0    &      \cite{Liang:2013ly}      &       \\ 
  110205A(g)     &    2.22 &    2.854  $ \pm$  0.145  &    53.748  $ \pm$  0.047  &     52.398  $ \pm$  0.06   &     2.91  &     \cite{Melandri:2014rc}        &       \\ 
  110213A(s)     &    1.46 &    2.382  $ \pm$  0.023  &    52.806  $ \pm$  0.041  &     52.32   $ \pm$  0.012  &     2.51  &       \cite{Cucchiara:2011wd}      &       \\ 
   110422A     &    1.77 &    2.624  $ \pm$  0.014  &    53.863  $ \pm$  0.042  &     53.446  $ \pm$  0.054  &     $\le$1.77  &   Gorbovskoy et al. 2011     &       \\ 
  110503A     &    1.61 &    2.757  $ \pm$  0.039  &    53.255  $ \pm$  0.048  &     53.255  $ \pm$  0.043  &     $\le$2.44  &   \cite{Melandri:2014rc}         &       \\ 
  110715A     &    0.82 &    2.339  $ \pm$  0.044  &    52.681  $ \pm$  0.045  &     52.622  $ \pm$  0.045  &     $\le$1.93  & \cite{Sanchez-Ramirez:2017nx}      &       \\ 
  110731(gL)      &    2.83 &    3.084  $ \pm$  0.014  &    53.602  $ \pm$  0.043  &     53.431  $ \pm$  0.048  &     0.7   &    \cite{Ackermann:2013fk}         &  \\
  111107A     &    2.89 &    2.623  $ \pm$  0.129  &    52.477  $ \pm$  0.072  &     52.236  $ \pm$  0.077  &     $\le$2.94  &    \cite{Lacluyze:2011lr}      &       \\ 
  111228A     &    0.71 &    1.766  $ \pm$  0.051  &    52.613  $ \pm$  0.048  &     51.766  $ \pm$  0.05   &     $\le$1.81  &     Melandri priv. com      &       \\ 
  120119A     &    1.73 &    2.62   $ \pm$  0.057  &    53.531  $ \pm$  0.064  &     52.9    $ \pm$  0.063  &     $\le$1.51  &   Melandri priv. com        &       \\ 
  120326A     &    1.8  &    2.061  $ \pm$  0.072  &    52.556  $ \pm$  0.072  &     52.017  $ \pm$  0.106  &    $\le$2.12  &     \cite{Melandri:2014eu}      &       \\ 
  120711A(g)     &    1.4  &    3.369  $ \pm$  0.016  &    54.176  $ \pm$  0.043  &     53.182  $ \pm$  0.015  &     2.38  &   \cite{Martin-Carrillo:2014oq}    &       \\ 
  120811C     &    2.67 &    2.297  $ \pm$  0.024  &    52.732  $ \pm$  0.018  &     52.35   $ \pm$  0.048  &     $\le$3.01  &     \cite{Denisenko:2012it}       &       \\ 
  120815A(s)     &    2.36 &    2.19   $ \pm$  0.087  &    52.086  $ \pm$  0.043  &     51.954  $ \pm$  0.043  &     2.64  &     \cite{Kruhler:2013gd}         &     \\ 
  120907A     &    0.97 &    2.484  $ \pm$  0.093  &    51.301  $ \pm$  0.065  &     51.38   $ \pm$  0.09   &     $\le$2.34  &     \cite{Oates:2012qe}     &       \\ 
  120909A(g)     &    3.93 &    3.261  $ \pm$  0.164  &    53.863  $ \pm$  0.042  &     52.857  $ \pm$  0.084  &     2.46  &     Melandri priv. comm.        &       \\ 
  120922A(g)     &    3.1  &    2.193  $ \pm$  0.045  &    53.301  $ \pm$  0.065  &     52.462  $ \pm$  0.06   &     2.95  &     \cite{Kuin:2012cs}       &       \\ 
  121128A(g)     &    2.2  &    2.297  $ \pm$  0.033  &    53.146  $ \pm$  0.031  &     52.806  $ \pm$  0.034  &     1.87  &     \cite{Wren:2012pi}         &       \\ 
  121211A     &    1.02 &    2.288  $ \pm$  0.058  &    51.114  $ \pm$  0.043  &     51.079  $ \pm$  0.054  &     $\le$2.35   &     \cite{Japelj:2012jt}      &       \\ 
  130215A(g)     &    0.6  &    2.394  $ \pm$  0.177  &    52.398  $ \pm$  0.035  &     50.924  $ \pm$  0.041  &     2.87  &     \cite{Zheng:2013hb}         &       \\ 
  130408A     &    3.76 &    3.002  $ \pm$  0.06   &    53.447  $ \pm$  0.082  &     53.763  $ \pm$  0.052  &     $\le$3.75  &     \cite{Trotter:2013sf}      &        \\ 
  130420A(g)     &    1.3  &    2.111  $ \pm$  0.024  &    52.857  $ \pm$  0.042  &     51.544  $ \pm$  0.037  &     2.551 &      \cite{Trotter:2013yf}       &       \\ 
  130427A(gL)     &    0.34 &    3.139  $ \pm$  0.0030 &    53.908  $ \pm$  0.043  &     53.431  $ \pm$  0.048  &     1.34  &    \cite{Ackermann:2014vl}         &       \\ 
  130505A     &    2.27 &    3.314  $ \pm$  0.021  &    54.748  $ \pm$  0.0090 &     54.602  $ \pm$  0.018  &     $\le$2.26  &   \cite{de-Pasquale:2013ad}         &       \\ 
  130606A     &    5.91 &    3.308  $ \pm$  0.103  &    53.431  $ \pm$  0.048  &     53.362  $ \pm$  0.038  &     $\le$3.43  &           \cite{Virgili:2013zp}   &       \\ 
  130610A(g)     &    2.09 &    2.96   $ \pm$  0.063  &    52.833  $ \pm$  0.045  &     52.38   $ \pm$  0.09   &     2.31  &   Trotter et al. (2013b)         &       \\ 
  130612A(g)     &    2.01 &    2.27   $ \pm$  0.075  &    51.855  $ \pm$  0.067  &     51.942  $ \pm$  0.045  &     2.04  &   Trotter et al. (2013c)         &       \\ 
  130701A     &    1.15 &    2.283  $ \pm$  0.02   &    52.431  $ \pm$  0.021  &     52.602  $ \pm$  0.033  &     $\le$2.44   &    \cite{Breeveld:2013fb}      &       \\ 
  130831A(g)     &    0.48 &    1.908  $ \pm$  0.032  &    51.879  $ \pm$  0.029  &     51.471  $ \pm$  0.054  &     2.86  &    \cite{Zhang:2016kl}     &       \\ 
  130907A     &    1.24 &    2.945  $ \pm$  0.012  &    54.502  $ \pm$  0.03   &     53.272  $ \pm$  0.019  &    $\le$1.64  &   \cite{Veres:2015ai}       &       \\ 
  131030A     &    1.29 &    2.593  $ \pm$  0.025  &    53.477  $ \pm$  0.029  &     53.0    $ \pm$  0.043  &     $\le$2.83  &     \cite{King:2014zm}       &       \\ 
  131108A     &    2.4  &    3.103  $ \pm$  0.016  &    53.778  $ \pm$  0.0070 &     53.301  $ \pm$  0.022  &     $\le$4.39  &     \cite{Gorosabel:2013ix}    &       \\ 
  131117A     &    4.04 &    2.346  $ \pm$  0.072  &    52.0    $ \pm$  0.087  &     51.919  $ \pm$  0.068  &     $\le$1.95  &     \cite{Trotter:2013rp}     &       \\ 
  131231A(g)     &    0.64 &    2.38   $ \pm$  0.0090 &    51.544  $ \pm$  0.0010 &     52.23   $ \pm$  0.0080 &     2.0    &     \cite{Liu:2014if}         &       \\ 
  140206A     &    2.73 &    3.044  $ \pm$  0.011  &    54.336  $ \pm$  0.0040 &     53.771  $ \pm$  0.01   &     $\le$2.477  &     \cite{Lien:2014ef}       &       \\ 
  140213A     &    1.21 &    2.344  $ \pm$  0.018  &    52.908  $ \pm$  0.027  &     52.663  $ \pm$  0.047  &     $\le$2.04  &   Trotter et al. (2014d)   &       \\ 
  140226A     &    1.98 &    3.091  $ \pm$  0.083  &    52.748  $ \pm$  0.085  &     52.405  $ \pm$  0.091  &     $\le$2.94  &   \cite{Cenko:2015cl}        &       \\ 
  140301A(g)     &    1.42 &    2.093  $ \pm$  0.035  &    51.623  $ \pm$  0.041  &     51.0    $ \pm$  0.043  &   $\le$4.25  &    \cite{Kruehler:2014kl}       &      \\ 
  140304A     &    5.28  &    3.067  $ \pm$  0.082  &    53.124  $ \pm$  0.033  &     52.944  $ \pm$  0.074  &   $\le$1.91  &    \cite{Gorbovskoy:2014rw}    &       \\ 
  140419A     &    3.96 &    3.162  $ \pm$  0.124  &    54.158  $ \pm$  0.151  &     53.756  $ \pm$  0.152  &     $\le$2.28    &     \cite{Zheng:2014uk}       &       \\ 
  140423A(g)     &    3.26 &    2.727  $ \pm$  0.031  &    53.748  $ \pm$  0.023  &     52.753  $ \pm$  0.049  &    2.30  &     \cite{Zheng:2014dp}    &       \\ 
  140506A     &    0.89 &    2.097  $ \pm$  0.236  &    52.041  $ \pm$  0.039  &     52.0    $ \pm$  0.087  &     $\le$2.26  &     \cite{Siegel:2014eq}       &       \\ 
  140508A      &    1.02 &    2.715  $ \pm$  0.02   &    53.358  $ \pm$  0.013  &     52.851  $ \pm$  0.024  &     $\le$4.38  &      \cite{Singer:2014fq}    &       \\ 
  140512A     &    0.73 &    3.006  $ \pm$  0.062  &    52.889  $ \pm$  0.022  &     51.729  $ \pm$  0.041  &     $\le$2.24  &      \cite{Gorbovskoy:2014qa}    &       \\ 
  140518A     &    4.71 &    2.4    $ \pm$  0.074  &    52.74   $ \pm$  0.103  &     52.279  $ \pm$  0.091  &     $\le$2.16  &   \cite{Trotter:2014oz}   &       \\ 
  140620A     &    2.04 &    2.371  $ \pm$  0.028  &    52.787  $ \pm$  0.021  &     52.176  $ \pm$  0.029  &   $\le$2.95  &    \cite{Kasliwal:2014wo}       &       \\ 
  140623A     &    1.92 &    3.009  $ \pm$  0.186  &    52.875  $ \pm$  0.052  &     51.851  $ \pm$  0.092  &     $\le$2.99  &    \cite{Bhalerao:2014si}        &       \\ 
  140629A(g)     &    2.28 &    2.45   $ \pm$  0.086  &    52.778  $ \pm$  0.043  &     52.431  $ \pm$  0.097  &     2.18  &    \cite{Gorbovskoy:2014sp}       &       \\ 
  140801A     &    1.32 &    2.4    $ \pm$  0.021  &    52.699  $ \pm$  0.017  &     52.724  $ \pm$  0.049  &   $\le$2.03  &  \cite{Lipunov:2016ez}         &       \\ 
  140808A     &    3.29 &    2.693  $ \pm$  0.03   &    52.886  $ \pm$  0.045  &     53.0    $ \pm$  0.043  &     $\le$4.07   &  Singer et al. (2014)          &       \\ 
  140907A     &    1.21 &    2.398  $ \pm$  0.026  &    52.352  $ \pm$  0.015  &     51.519  $ \pm$  0.039  &     $\le$3.74  &     \cite{Volnova:2014tx}    &       \\ 
  141028A     &    2.33 &    2.921  $ \pm$  0.022  &    53.778  $ \pm$  0.043  &     53.27   $ \pm$  0.047  &     $\le$4.57  &  \cite{Burgess:2016ij}         &       \\ 
  141109A(g)     &    2.93 &    2.875  $ \pm$  0.136  &    53.491  $ \pm$  0.042  &     52.623  $ \pm$  0.052  &     2.98  &     Klotz et al. 2014        &       \\ 
  141220A     &    1.32 &    2.433  $ \pm$  0.167  &    52.255  $ \pm$  0.048  &     52.072  $ \pm$  0.074  &     $\le$2.68  &   \cite{Gorosabel:2014ng}         &       \\ 
  141221A(g)     &    1.45 &    2.571  $ \pm$  0.083  &    52.279  $ \pm$  0.046  &     51.845  $ \pm$  0.056  &     2.04  &     \cite{Bardho:2016bq}         &       \\ 
  141225A     &    0.92 &    2.554  $ \pm$  0.068  &    52.326  $ \pm$  0.041  &     51.477  $ \pm$  0.072  &    $\le$2.32  & \cite{Buckley:2014th}        &       \\ 
  150206A     &    2.09 &    2.848  $ \pm$  0.068  &    53.699  $ \pm$  0.043  &     53.352  $ \pm$  0.058  &     $\le$2.72  &  \cite{Oates:2015bs}     &       \\ 
  150301B     &    1.52 &    2.809  $ \pm$  0.055  &    52.45   $ \pm$  0.046  &     51.881  $ \pm$  0.046  &     $\le$1.90  &   \cite{Gorbovskoy:2016qc}         &       \\ 
  150314A     &    1.76 &    2.933  $ \pm$  0.0070 &    54.255  $ \pm$  0.048  &     53.914  $ \pm$  0.011  &     $\le$2.13  &    \cite{Zheng:2015xr}       &       \\ 
  150323A     &    0.59 &    2.179  $ \pm$  0.069  &    52.114  $ \pm$  0.067  &     50.352  $ \pm$  0.097  &     $\le$2.68  &      \cite{Cenko:2015tk}      &       \\ 
  150514A     &    0.81 &    2.121  $ \pm$  0.036  &    52.041  $ \pm$  0.039  &     51.778  $ \pm$  0.029  &     $\le$4.62   &     \cite{Marshall:2015hh}       &       \\ 
  150818A     &    0.28 &    2.107  $ \pm$  0.102  &    51.079  $ \pm$  0.036  &     49.724  $ \pm$  0.09   &      $\le$4.14  &   \cite{Mazaeva:2015qr}     &       \\ 
  150821A     &    0.75 &    2.788  $ \pm$  0.203  &    53.164  $ \pm$  0.089  &     53.0    $ \pm$  0.03   &     $\le$2.50   &    \cite{Kuin:2015qd}      &       \\ 
  151021A     &    2.33 &    2.753  $ \pm$  0.051  &    54.0    $ \pm$  0.043  &     53.322  $ \pm$  0.124  &     $\le$2.15  &  \cite{Trotter:2015pi}       &       \\ 
  160509A     &    1.17 &    2.796  $ \pm$  0.069  &    53.959  $ \pm$  0.043  &     53.301  $ \pm$  0.033  &     $\le$4.33  &      \cite{Laskar:2016fy}     &       \\ 
  160629A(g)     &    3.33 &    3.108  $ \pm$  0.028  &    53.672  $ \pm$  0.028  &     52.959  $ \pm$  0.076  &     1.91  &  \cite{Klotz:2016ec}       &       \\ 
\hline\hline
\caption{Sample of GRBs with measured \tp\ (69 events) from the peak of the light curve (those from the LAT light curve are labelled ``L"). (g) = Gold sample, (s) = Silver sample. Upper limits on \tp\ for 106 GRBs are reported.
}
\label{tabTOT}
\end{longtable}

\noindent
\twocolumn

\section*{Acknowledgments}
INAF PRIN 2014 (1.05.01.94.12) is acknowledged. 
The anonymous referee is acknowledged for useful comments that helped to improve the content of the paper.

\bibliographystyle{aa}
\bibliography{references}

\end{document}